\documentclass[smallextended]{svjour3}
 \usepackage{amsmath,amssymb,amsbsy,amstext,amscd,amsfonts}
 \usepackage{exscale,calrsfs,times,theorem}
 \usepackage{color}
\newcommand{\al}{\alpha}

\newcommand{\vf}{\varphi}

\newcommand{\cO}{{\mathcal O}}

\newcommand{\bC}{\mathbb{C}}
\newcommand{\bH}{\mathbb{H}}
\newcommand{\bL}{\mathbb{L}}

\newcommand{\bN}{\mathbb{N}}
\newcommand{\bR}{\mathbb{R}}
\newcommand{\bS}{\mathbb{S}}
\newcommand{\bV}{\mathbb{V}}
\newcommand{\bW}{\mathbb{W}}

\newcommand{\bZ}{\mathbb{Z}}
\newcommand{\cA}{{\mathcal A}}
\newcommand{\cB}{{\mathcal B}}

\newcommand{\cF}{{\mathcal F}}

\newcommand{\cK}{{\mathcal K}}
\newcommand{\cL}{{\mathcal L}}
\newcommand{\cM}{{\mathcal M}}

\newcommand{\cP}{{\mathcal P}}

\newcommand{\cT}{{\mathcal T}}

\newcommand{\fB}{\mathfrak{B}}

\newcommand{\fD}{\mathfrak{D}}

\newcommand{\fM}{\mathfrak{M}}

\newcommand{\dst}{\displaystyle}
\newcommand{\pa}{\partial}
\newcommand{\ov}{\overline}
\newcommand{\wt}{\widetilde}

\newcommand{\A}{{\boldsymbol A}}
\newcommand{\B}{{\boldsymbol B}}
\newcommand{\D}{{\boldsymbol D}}
\newcommand{\F}{{\boldsymbol F}}

\newcommand{\Kb}{{\boldsymbol K}}

\newcommand{\Lbb}{{\boldsymbol \Lambda}}
\newcommand{\K}{{\boldsymbol K}}
\newcommand{\Lb}{{\boldsymbol L}}

\newcommand{\Pb}{{\boldsymbol P}}
\newcommand{\R}{{\boldsymbol R}}
\newcommand{\Sb}{{\boldsymbol S}}
\newcommand{\T}{{\boldsymbol T}}
\newcommand{\V}{{\boldsymbol V}}

\newcommand{\scal}{\mbox{\bf{(}}}
\newcommand{\scar}{\mbox{\bf{)}}}

\newcommand{{\Div}}{{\bf\rm div}}
\newcommand{\sign}{\operatorname{sign}}

\newcommand{\QED}{\hspace{\fill}$\Box$\medskip\par}

\newcommand{\la}{\langle}
\newcommand{\ra}{\rangle}

\begin{document}

\title{CONVOLUTION EQUATIONS ON THE LIE GROUP $G=(-1,1)$}

\titlerunning{CONVOLUTION EQUATIONS ON THE LIE GROUP}

\author{Roland Duduchava.}

\institute{R. Duduchava\at The University of Georgia, Institute of Mathematics, IV, 77a Merab Kostava St, Tbilisi 0171, Georgia.\\
Tel.: +995-555333855\\
\email{RolDud@gmail.com} \at A.Razmadze Mathematical Institute, Tbilisi State University, Tamarashvili str. 6, Tbilisi 0177, Georgia}

\date{Received: date / Accepted: date}

\maketitle

 \begin{abstract}
 The interval $G=(-1,1)$ turns into a Lie group under the group operation $x\circ  y:=(x+y)(1+xy)^{-1},\qquad x,y\in G$. This enables definition of the invariant measure $dG(x):=(1-x^2)^{-1}dx$ and the Fourier transformation $\cF\hskip-1mm_G$ on the interval $G$ and, as a consequence, we can consider Fourier convolution operators $W^0_{G,a}:=\cF\hskip-1mm_G^{-1} a\cF\hskip-1mm_G$ on $G$. This class of convolutions includes celebrated Prandtl, Tricomi and Lavrentjev-Bitsadze equations and, also, differential equations of arbitrary order with the natural weighted derivative $\fD_G u(x)=(1-x^2)u'(x)$, $x\in G$. Equations are solved in the scale of Bessel potential $\bH^s_p(G,dG(x))$, $1\leqslant p\leqslant\infty$,  and H\"older-Zygmound  $\bZ^\nu(G)$, $0<\mu,\nu<\infty$ spaces, adapted to the group $G$. Boundedness of convolution operators (the problem of multipliers) is discussed. The symbol $a(\xi)$, $\xi\in\bR$, of a convolution equation $W^0_{G,a}u=f$ defines solvability: the equation is uniquely solvable if and only if the symbol $a$ is elliptic. The solution is written explicitly with the help of the inverse symbol.

 We touch shortly the multidimensional analogue-the Lie group $G^n$.
 \end{abstract}


\tableofcontents

\section*{Introduction}

The present investigation is inspired by papers of V.E. Petrov \cite{Pe06a,Pe06b,SP20}, where the author applied the finite interval Fourier transformation
 \begin{eqnarray}\label{e0.1}
(\cF\hskip-1mm_Gv)(\xi):=\int_{-1}^1\left(\frac{1+y}{1-y}\right)^{i\xi}
     \frac{v(y)dy}{1-y^2},\qquad \xi\in\bR
 \end{eqnarray}
to the investigation of convolution equations of the following type
 \begin{eqnarray}\label{e0.2}
c_0u(x)+\int_{-1}^1k\left(\frac{x-y}{1-xy}\right)
     \frac{v(y)dy}{1-y^2}=h(x),\qquad x\in G:=(-1,1).
 \end{eqnarray}
\eqref{e0.2} becomes convolution if the interval $G=(-1,1)$ is endowed with the group operation $x\circ y:=\dst\frac{x+y}{1+xy}$, making $G$ a Lie group (see details in \S\,1 below).

To the class of convolution equations \eqref{e0.2} belong the celebrated Prandtl equation
\begin{eqnarray}\label{e0.3}
\Pb u(x)=\frac{c_0u(x)}{1-x^2}+\frac{c_1}{\pi}\int_{-1}^1
     \dst\frac{u'(y)dy}{y-x}=f(x), \qquad x\in G
 \end{eqnarray}
(actually a special case of the Prandtl equation for a parabolic wing chord), singular Tricomi equation
\begin{eqnarray}\label{e0.4}
\T v(x)=c_0v(x)+\frac{c_1}\pi\int_{-1}^1\dst\frac{v(y)dy}{y-x}+\frac{c_2}\pi
     \int_{-1}^1\dst\frac{v(y)dy}{1-xy}=g(x), \qquad x\in (-1,1)
 \end{eqnarray}
and Lavrentjev-Bitsadze equation
\begin{eqnarray}\label{e0.5}
\Lb\B \varphi(x)=c_0\varphi(x)+\frac{c_1}\pi\int_0^1\left[\dst\frac1{y-x}+
     \dst\frac{1-2y}{x+y-2xy}\right]\varphi(y)dy=h(x),\\
x\in G^+:=(0,1). \nonumber
 \end{eqnarray}

Equations \eqref{e0.2}-\eqref{e0.4} and equation \eqref{e0.5} after the variable transformation mapping $G^+:=(0,1)\to G$ (see \eqref{e7.1} below), are particular cases of the following integro-differential equation
\begin{eqnarray}\label{e0.6}
\A u(x)&:=&\sum_{k=0}^m\left[c_k\fD^k_G u(x)+d_k\fD^{m_k}_G\int_{-1}^1
     \cK_k\left(\frac{x-y}{1-xy}\right)(\fD^{n_k}_G u)(y)\frac{dy}{1-y^2}\right]\nonumber\\
     &=& w(x),\qquad  x\in G=(-1,1),
\end{eqnarray}
where $c_1,\ldots,c_m$ and $d_1,\ldots,d_m$ are complex valued constants, $m_0,\ldots,m_m$, $n_0,\ldots,n_m$ are non-negative integers, $\cK_1,\ldots,\cK_m\in\bL_1(G,dG(x))$.  $\A=W^0_{G.a}$ represents a convolution operator on the Lie group $G$ (see \S\,\ref{s3} below) and
\begin{eqnarray}\label{e0.8}
\fD_G:=(1-x^2)\frac{d}{dx}.
 \end{eqnarray}
is the natural differential operator on this group, which means that it is a convolution operator and generates the Lie algebra on the Lie group $G$.

Equations \eqref{e0.2}--\eqref{e0.6}  have ample of applications in Mechanics and Mathematical physics and were investigated by many authors (see surveys in \cite{Pe06a,SP20}, \cite[\S\,20]{Du79}, \cite{Ka75} and the recent papers \cite{AA21,AP16}).

Equations \eqref{e0.3} and \eqref{e0.4}  were solved  by V. E. Petrov in \cite{Pe06a} (also see \cite{Pe06b,SP20}) by using the $\cP$-transformation, which he defined as the equivalent transformation $\cP=t_*\cF x_*$ to the classical Fourier transformation $\cF$ on the real axes under the transformations $t_*\vf(x):=\vf(t(x))$, $x_*\psi(t):=\psi(x(t))$ which are inverse to each-other. Here
\begin{eqnarray}\label{e0.7}
t(x)=\dst\frac12\ln\frac{1+x}{1-x},\qquad x(t)=\tanh\,t ,\qquad x\in G,\quad t \in\bR.
 \end{eqnarray}

We have noted that the transformation $\cP$ on a segment $G=(-1,1)$ considered in \cite{Pe06a} can be interpreted as the Fourier transformation on the group $G$. Then the equations  \eqref{e0.2}--\eqref{e0.5} are interpreted as convolutions and the Fourier transformation \eqref{e0.1}  applied to these equations transforms them to a simple operators of multiplication by the "symbol" (the Fourier image of the kernel) of the Fourier image of an unknown function. That motivated description of multipliers and finding criteria of invertibility of convolution operators. Also that motivated introduction of the appropriate Bessel potential spaces, based on the differential operator $\fD_G$ to consider these equations in better space settings.

These aspects were missed in \cite{Pe06a,Pe06b} and in the subsequent papers. Equations \eqref{e0.2}--\eqref{e0.4} and some boundary value problem for the equation \eqref{e0.5} on the part of the unit sphere, were solved in \cite{Pe06a,Pe06b} in the general Banach spaceless setting, while in \cite{SP20} equation \eqref{e0.3} was investigated in the Bessel potential space setting $\wt{\bH}^s(G)$, $-1\leqslant s\leqslant1$, which was defined as the image of the corresponding Bessel potential space on the axes $\wt{\bH}^s(G):=t_*\bH^s(\bR)$ under the isomorphism $t _*\varphi(x):=\varphi(t (x))$ (cf. \eqref{e0.7}).

In our approach we use a natural Fourier transformation on the Lie Group $\cF\hskip-1mm_G$ (cf. \S\,1 below), the natural differential operator $\fD_G$ (see \eqref{e0.1} and \eqref{e0.8},
and the Bessel potential operators $\Lbb^s=\cF\hskip-1mm_G^{-1} (1+|\xi|^2)^{s/2}\cF\hskip-1mm_G$, and define Bessel potential $\bH^s_p(G,dG(x))$ and Sobolev $\bW^m_p(G,dG(x))$ spaces for arbitrary $s\in\bR$, $1\leqslant p\leqslant\infty$, $m=0,1,\ldots$ in a standard way. A convolution operator $W^0_{G,a}:=\cF\hskip-1mm_G^{-1}a\cF\hskip-1mm_G$ on $G$ is defined and theorems on multipliers in the Bessel potential spaces are proved. Criteria for the Fredholm property and unique solvability of equations \eqref{e0.2}-\eqref{e0.6} in the Bessel potential spaces are found and explicit formulae for their solution  are indicated (see \S\,\S\, 4-7). H\"older-Zygmound spaces $\bZ^\nu(G)$, $0<\mu<\infty$ are defined and used to establish a priori smoothness of solutions to  equations \eqref{e0.2}--\eqref{e0.6}.

What we expose here for the group $G $ can also be done for the multidimensinal case $G^n:=G\times\cdots\times G$: Fourier transformations ${\cF\hskip-1mm_{G^n}}\!\!^{\pm1}$, Bessel potential $\bH^s_p(G^n,dG(x))$, Sobolev $\bW^m_p(G^n,dG(x))$ and H\"older-Zygmound $\bZ^\mu(G^n)$ spaces, theorems on multipliers and solvability results for multidimensional convolution equations $W^0_{G^n,a}U:={\cF\hskip-1mm_{G^n}}\!\!^{-1}a\cF\hskip-1mm_{G^n}U=F$, $U\in\bH^s_p(G^n,dG^n(x))$, $F\in\bH^{s-r}_p(G^n,dG(x))$ and the multi-variable symbol belongs to the multiplier class $a\in\cM_p(G^n)$ (see \S\, 3). Therefore some assertions will be formulated for multi-variable case (see, e.g., Proposition \ref{p3.7}).

The paper is organized as follows. In \S\, 1 we expose results on the Lie Group $G =(-1,1)$-invariant Haar measure, characters, Fourier transformation and its inverse, define  Lebesgue space $\bL_p(G,dG(x))$. In \S\, 2 we define Bessel potential $\bH^s_p(G,dG(x))$, Sobolev $\bW^m_p(G,dG(x))$ and H\"older-Zygmound $\bZ^\mu(G)$ spaces ($1\leqslant p\leqslant\infty$, $s\in\bR$, $m=1,2,\ldots$, $\mu>0$, $\gamma\in\bR$). In \S\, 3 convolution operators on the Lie group $G $ are defined and a couple of theorems on their boundedness are proved. In \S\, 4 -\S\, 7 we expose results on equations \eqref{e0.2}-\eqref{e0.5}.

\section{\bf Lie groups $G$ and $G^n$}

On the interval $G:=(-1,1)$ we define the following additional group operation
 \begin{eqnarray}\label{e1.1}
x\circ  y:=\frac{x+y}{1+xy},\qquad x,y\in G,
 \end{eqnarray}
which makes $G$ a Lie group. The inverse element to $x$ is $-x$ and the group operation is the isomorphism of the Lie group $x\circ   y:G \times G \rightarrow G .$ Indeed,
\[
\partial_y\dst\frac{x+y}{1+xy}=\dst\frac{1+xy-x(x+y)}{(1+xy)^2}=\dst\frac{1-x^2}{(1+xy)^2}>0;
\]
therefore, the function $\dst\frac{x+y}{1+xy}$ is increasing with respect of both variables $x,y\in G$ and
\begin{eqnarray*}
\inf_{x,y\in G}\dst\frac{x+y}{1+xy}=\dst\frac{-1-1}{1+1}=-1\\
\sup_{x,y\in G}\dst\frac{x+y}{1+xy}=\dst\frac{1+1}{1+1}=1.
\end{eqnarray*}
This accomplishes the proof that the binary operation $x\circ  y$ makes $G $ a Lie group.

The invariant measure is
 \[
dG(x):=\frac{dx}{1-x^2}.
 \]
Indeed, we have
 \begin{eqnarray}\label{e1.2}
dG(x\circ  y)=\frac{\dst d_x\frac{x+y}{1+xy}}{1-\left(\dst\frac{x+y}{1+xy}\right)^2}
     =\frac{1-y^2}{(1-x^2)(1-y^2)}dx=\frac{dx}{1-x^2}=dG(x).
 \end{eqnarray}

Characters of the Lie group $G $ coincides with the set of isomorpisms to the unit circle and are given by the mappings
 \[
x\longrightarrow C(x,\xi):=\left(\frac{1+x}{1-x}\right)^{i\xi}\qquad\forall\,x\in G.
 \]
Indeed, to prove this we have to check that $C(x\circ  y,\xi)=C(x,\xi)C(y,\xi),\quad \forall\,x,y\in G $. Indeed,
 \begin{eqnarray}\label{e1.3}
&&C(x\circ  y,\xi)=\left(\frac{1+(x\circ y)}{1=(x\circ y)}
     \right)^{i\xi}=\left(\frac{1+\dst\frac{x+y}{1+xy}}{1
     -\dst\frac{x+y}{1+xy}}\right)^{i\xi}
     =\left(\frac{1+xy+x+y}{1+xy-x-y}\right)^{i\xi}\nonumber\\
&&\hskip10mm=\left(\frac{1+x}{1-x}\right)^{i\xi}\left(\frac{1+y}{1
     -y}\right)^{i\xi}=C(x,\xi)C(y,\xi)\qquad \forall\,x,y\in G .
 \end{eqnarray}
Thus, the Fourier transformation on the space of functions on the group $G $ is defined as follows (cf. \eqref{e0.1}):
 \begin{eqnarray}\label{e1.4}
(\cF\hskip-1mm_Gv)(\xi):=\int_{-1}^1C(y,\xi)v(y)dG(y)
      =\int_{-1}^1\left(\frac{1+y}{1-y}\right)^{i\xi}\frac{v(y)dy}{1-y^2}.
 \end{eqnarray}

The pull back operators, corresponding to the diffeomorphisms \eqref{e0.7}  of the Lie groups $\bR$ and $G$ represent isometric isomorphisms
\begin{eqnarray}\label{e1.5}
\begin{array}{c}
(t_*\varphi)(x):=\varphi(t (x))=\varphi\left(\dst\frac12\ln\frac{1+x}{1-x}\right)\;:\;
     \mathbb{L}_p(\mathbb{R})\rightarrow\mathbb{L}_p(G,dG(x)),\\[3mm]
(x_*\varphi_0)(t):=\varphi_0(x(t ))=\varphi_0(\tanh\,t ):\mathbb{L}_p(G,dG(x))\rightarrow
     \mathbb{L}_p(\mathbb{R}),
\end{array}\\
     x_*(t)=t^{-1}_*(t), \qquad t_*(x)=x^{-1}_*(x),\qquad x\in G,\qquad t \in\mathbb{R} \nonumber
\end{eqnarray}
of the Lebesgue spaces $\mathbb{L}_p(\mathbb{R})$ and $\mathbb{L}_p(G,dG(x))$, $1\leqslant p\leqslant\infty$, where the latter weighted Lebesgue space is equipped with the norm
\begin{eqnarray}\label{e1.6}
\begin{array}{l}
\|\varphi\big|\bL_p(G,dG(x))\|:=\left[\dst\int_{-1}^1|\vf(y)|^p
     dG(y)\right]^{1/p}=\left[\dst\int_{-1}^1|\vf(y)|^p\dst\frac{dy}
     {1-y^2}\right]^{1/p}\\[4mm]
\hskip70mm\quad \mbox{for}\quad 1\leqslant p<\infty\\[2mm]
\|\vf\big|\mathbb{L}_\infty(G,dG(x))\|:=\|\vf\big|\mathbb{L}_\infty
     (G)\|={\rm ess}\,\sup_{x\in G}|\varphi(x)| \quad \mbox{for}
     \quad p=\infty.
\end{array}
\end{eqnarray}

The dual space to $\bL_p(G,dG(x))$, with respect to the scalar product
 \[
\scal\vf,\psi\scar_G:=\dst\int_{-1}^1\vf(y)\ov{\psi(y)}
     \dst\frac{dy}{1-y^2}, \quad 1<p<\infty,
 \]
is the space $\bL_{p'}(G,dG(x))$, $p':=\dst\frac p{p-1}$.

Using the transformation of the variables \eqref{e0.7}  and taking into account the connection between differentials
 \[
dt =\dst\frac{1}{2}d\left[\ln\frac{1+x}{1-x}\right]
     =\dst\frac{(1-x)+(1+x)}{2\dst\frac{1+x}{1-x}(1-x)^2}dx
     =\dst\frac{dx}{1-x^2}
 \]
we easily prove the following equality between the norms (isometrical isomorphism of the spaces)
\begin{eqnarray}\label{e1.7}
&&\|\varphi_0\big|\mathbb{L}_p(G, dG(x))\|:=\left[\int_{-1}^{1}|\varphi_0(x)|^p
     \,dG(x)\right]^{1/p}\\
&&=\left[\int_{-1}^{1}\left|\varphi\left(\dst\frac12
     \ln\frac{1+x}{1-x}\right)\right|^p\,\dst\frac{dx}{1-x^2}\right]^{1/p}
     =\left[\int_{-\infty}^\infty|\varphi(t)|^p\,dt\right]^{1/p}
     =\|\varphi\big|\mathbb{L}_p(\bR)\|,\nonumber
\end{eqnarray}
where we used the substitution $t=\dst\frac12\ln\dst\frac{1+x}{1-x}$, $dt=\dst\frac{dx}{1-x^2}$ and
 \[
\varphi\in\bL_p(\bR),\qquad \varphi_0(x):=t _*\varphi(x)=\varphi\left(\dst\frac12
     \ln\frac{1+x}{1-x}\right),\quad\varphi_0\in\mathbb{L}_p(G,dG(x)).
 \]

The Fourier transform $\cF\hskip-1mm_G\vf_0(\xi)$ of a function $\vf_0(x)$ on the Lie group $G$ and the Fourier transform $\cF_\bR\vf(\xi)$ of a function $\vf(t)$ on the Lie group $\bR$ are related by the formulae
 \begin{eqnarray}\label{e1.8}
\hskip-3mm
 \begin{array}{l}
(\cF\hskip-1mm_G\varphi_0)(\xi)=\D_2\cF_\bR( x_*\varphi_0)(\xi)
     =\cF_\bR(x_*\varphi_0)(2\xi),\qquad\xi\in\bR,\\[3mm]
({\cF}_G^{-1}\psi)(x)=(t _*\cF^{-1}_\bR\D_{1/2}\psi)(x)
     =\dst\frac1{\pi}\dst\int_{-\infty}^\infty\left(\frac{1+x}{1-x}
     \right)^{-i\xi}\hskip-5mm\psi(\xi/2)d\xi,\quad x\in G,\\[3mm]
\hskip50mm\varphi_0\in C^\infty_0(G),\qquad \psi\in C^\infty_0(\bR),
 \end{array}
 \end{eqnarray}
where $\D_\lambda\psi(\xi):=\psi(\lambda\xi)$ is the dilation operator and $x_*$ is defined in \eqref{e1.5}. $C^\infty_0(\bR)$ and $C^\infty_0(G)$ denote the space of $C^\infty-$functions with compact supports on the sets $G=(-1,1)$ and $\bR$.

The isomorphism of spaces \eqref{e1.5} and \eqref{e1.7} via transformation \eqref{e0.7}  was noted in \cite{Pe06a} (also see \cite{SP20}) and applied to the definition of the transformation \eqref{e1.4}, which was called there $\cP$-transformation.

The following Parseval's equality is valid:
 \begin{eqnarray}\label{e1.9}
\scal\cF\hskip-1mm_G\vf,\cF\hskip-1mm_G\psi\scar_\bR&=&
     \int_{-\infty}^\infty\hskip-3mm\cF\hskip-1mm_G\vf(\xi)\ov{\cF\hskip-1mm_G
     \psi(\xi)}d\xi=\int_{-\infty}^\infty\cF\hskip-1mm_G\vf(\xi)
     \int_{-1}^1\left(\frac{1+y}{1-y}\right)^{-i\xi}
     \frac{\ov{\psi(y)}dy}{1-y^2}d\xi\nonumber\\
&=&\pi\int_{-1}^1\ov{\psi(y)}\frac1\pi\int_{-\infty}^\infty
     \left(\frac{1+y}{1-y}\right)^{-i\xi}\cF\hskip-1mm_G
     \vf(\xi)d\xi\frac{dy}{1-y^2}\nonumber\\
&=&\pi\int_{-1}^1\ov{\psi(y)}\cF\hskip-1mm_G^{-1}\cF\hskip-1mm_G
     \vf(y)dG(y)=\pi\int_{-1}^1\vf(y)\ov{\psi(y)}dG(y)\nonumber\\
&=&\pi\scal\vf,\psi\scar_G.
 \end{eqnarray}
From \eqref{e1.9} follows that the mappings
 \begin{eqnarray}\label{e1.10}
 \begin{array}{l}
\dst\frac1{\sqrt{\pi}}\cF\hskip-1mm_G\; :\;\bL_2(G,dG(x))\to\bL_2(\bR),\\[3mm] \sqrt{\pi}
{\cF}_G^{-1}\;:\;\bL_2(\bR)\to\bL_2(G,dG(x))
 \end{array}
\end{eqnarray}
are isometric isomorphisms of the spaces.

Note, that the direct product $G^n:=G\times\ldots\times G$ of $n$ copies of intervals $G:=(-1,1)$ endowed with the group operation
 \begin{eqnarray*}
&x\circ y:=\left(x_1\circ y_1,\ldots,x_n\circ y_n\right),\\
&x=(x_1,\ldots,x_n)^\top,y=(y_1,\ldots,y_n)^\top\in G^n,\nonumber
 \end{eqnarray*}
makes $G^n$ a Lie group. The inverse element to $x$ is $-x$ and the invariant measure is
 \[
dG(x):=dG(y_1)\cdots dG(y_n)=\frac{dx_1}{1-x^2_1} \cdots\frac{dx_n}{1-x^2_n},\quad x=(x_1,\ldots,x_n)\in G^n.
 \]
The Fourier transformation on the space of functions on the group $G^n$ is:
 \begin{eqnarray*}
(\cF\hskip-1mm_G v)(\xi):=\int_{G^n}\left(\frac{1+y}{1-y}
     \right)^{i\xi}v(y)dG(y):=\int_{G^n}\prod_{k=1}^n\left(\frac{1
     +y_k}{1-y_k} \right)^{i\xi_k}v(y)dG(y),\\
\xi\in\bR^n.
 \end{eqnarray*}

Further we will formulate and prove assertions for the group $G $, although all of them from \S\S\, 2-3, except the second part of Theorem \ref{t3.2}, are valid for the group $G^n$.

\section{\bf Function spaces on  $G$}

Let $\bS(\bR)$ denote the Schwartz space of fast decaying $C^\infty(\bR)-$functions, which eliminate at $\pm\infty$ together with the derivatives $[(1+t^2)]^m(d^k\vf)(t)/dt^k$ for arbitrary integers $m,k=0,1,\ldots$. $\bS(G)$ denotes the image of $\bS(\bR)$ under the transformation (cf. \eqref{e1.5})
\begin{eqnarray}\label{e19}
t_*\;:\;\bS(\bR)\to\bS(G)
\end{eqnarray}
and consists of $C^\infty(G)-$functions $\vf(x)$ which eliminate at $\pm1$ together with derivatives $[\ln(1-x^2)]^m(\fD^k\vf)(x)$ for arbitrary integers $m,k=0,1,\ldots$. The notations $\bS^\prime(\bR)$ and $\bS^\prime(G)$ are used for the dual Schwartz space of distributions. It is well known that the spaces
$\bS(\bR)$ and $\bS^\prime(\bR)$ are invariant under the Fourier transformations $\cF_\bR^{\pm1}$ and, therefore, due to \eqref{e1.8}, \eqref{e19}, the Fourier transformations $\cF\hskip-1mm_G^{\pm1}$ map the spaces as follows:
\begin{eqnarray}\label{e20}
 \begin{array}{rcl}
\cF\hskip-1mm_G&:&\bS(G)\to\bS(\bR),\\[1mm]
     &:&\bS^\prime(G)\to\bS^\prime(\bR).
 \end{array}\hskip10mm
 \begin{array}{rcl}
\cF\hskip-1mm_G^{-1}&:&\bS(\bR)\to\bS(G),\\[1mm]
     &:&\bS^\prime(\bR)\to\bS^\prime(G).
 \end{array}
\end{eqnarray}

Let $W^0_a$ denote the Fourier convolution operator on the real axes
 \[
W^0_a\vf:=\cF^{-1}_{\bR}a\cF_{\bR}\vf, \qquad \vf\in\bS(\bR)
 \]
and $a(\xi)$ is called its symbol (cf. \cite{Du79}).

The notation $\bH^m_p(\bR)=\bW_p^m(\bR)$ with $1\leqslant p\leqslant\infty,\; m\in \bN_0$ refers to the Sobolev space of which represents the closure of the Schwartz space $\bS(\bR)$ under the norm
 \begin{eqnarray*}
\| f\,|\, \bW_p^m(\bR)\|\,:=\,\left(\sum\limits_{k=0}^m\|\partial^k_t  f \,|\,
    \bL_p(\bR)\|^p\right)^{1/p}
 \end{eqnarray*}
for $1\leqslant p<\infty$, with the usual ${\rm ess}\,\sup$-norm modification for $p=\infty$:
 \[
\|f\,|\, \bW_\infty^m(\bR)\|\,:=\,\sum\limits_{k=0}^m{\rm ess}\sup\limits_{t \in\bR}|\pa^k_t  f(t )|\, .
 \]

The notation $\bH^s_p(\bR)$  with $s\in\bR,\;1\leqslant p\leqslant\infty$ refers to the Bessel potential space on the Lie group $\bR$, which represents the closure of the Schwartz space $\bS(\bR)$ under the norm
 \begin{eqnarray*}
\|f\,|\,\bH_p^s(\bR)\| \,:=\, \| \cF^{-1}_{\bR} \langle\xi
     \rangle^s \cF_{\bR}f \,|\, \bL_p(\bR)\|=\, \| W^0_{\la\cdot\ra^s}f\,|\, \bL_p(\bR)\|<+\infty, \\
 \langle\xi\rangle^s:=(1+|\xi|^2)^{s/2}\nonumber
 \end{eqnarray*}
for $1\leqslant p\leqslant\infty$. We will apply the standard convention and write $\bH^s(G)$ for the Hilbert space $\bH^s_2(G)$, dropping the subscript index $p=2$. Due to the classical Parseval's equality for the axes $\bR$ the formula
 \begin{eqnarray}\label{e2.2}
\|f\,|\,\bH^s(\bR)\|_0:=\left(\int_{\bR}\left|(1+\xi^2)^{s/2}
    \cF_{\bR}f(\xi)\right|^2\,d\xi\right)^{1/2}
 \end{eqnarray}
provides an equivalent norm in the Hilbert space $\bH^s(\bR)$.

If $\langle\xi\rangle^{-r}a(\xi)$ has bounded variation on $\bR$, than $W^0_a\;:\;\bH^s_p(\bR)\to\bH^{s-r}_p(\bR)$ extends to a bounded operator for arbitrary $s,r\in\bR$, $1<p<\infty$ (see, e.g., \cite{Du79}).

It is well known that the derivative $\pa_t\vf(t)=\varphi^\prime(t)=(W^0_{-i\xi}\vf)(t)$ on the real axes $\bR$ is a convolution operator and its symbol is $-i\xi$:
\begin{eqnarray}\label{e2.3}
(\cF_{\bR}\pa_t\varphi)(\xi)=-i\xi(\cF_\bR\varphi)(\xi),\qquad \xi\in\bR.
\end{eqnarray}

Let $ 0<\alpha<1,\; \; \; 1\leqslant p\leqslant\infty{}.$ Then the space $ \bZ_p^\alpha(\bR)$ consists of functions
 \[
\bZ_p^\alpha(\bR)=\left\{\vf\in\bL_p(\bR)\; \; : \; \; \| \vf \| _{\bZ^{\alpha{}}}=\sup_{x,t\in\R}\frac{\mid \vf (x)-\vf(t)\mid }{\mid x-t\mid^\alpha}<\infty\right\}
 \]
and is endowed with the norm
\begin{equation}\label{e2.3a}
\|\vf\|_{\bZ_{p}^{\alpha{}}}=\|\vf\|_{\bL_{p}} + \|\vf\| _{\bZ^{\alpha{}}}.
\end{equation}

To extend the definition of the space $\bZ_p^\alpha(\bR)$ to the case $\alpha\geqslant1$ the Poisson integral is involved (cf. \cite[\S\, III.2]{St70}
\begin{eqnarray*}
P_{y}\vf (x) = \int_\bR P_{y}(x-t)\vf(t)dt = W^0_{a_{y}}f(x),
     \hspace{1cm} P_{y}(x)=\frac1{\pi}\frac{y}{x^2 + y^2}, \\
a_{y}(\xi )=\exp (-\mid \xi \mid y), \hspace{5mm} y>0, \; \; \; x,\xi\in\bR.
\end{eqnarray*}

$P_y\vf(x)$ approximates $\vf(x)$ (cf. \cite[\S\, III.2]{St70})
\[
\lim_{y\ra 0}\| P_{y}\vf - \vf \| _{p}=0, \hspace{1cm} \vf \in \bL_p(\bR),\qquad 1<p<\infty{}.
\]
\begin{lemma}
For $ 0<\alpha<1$, $1\leqslant p\leqslant\infty$ the expressions
\begin{eqnarray*}
&\|\vf\| ^{(0)}_{\bZ^{\alpha{}}_{p}}=\| \vf \|_{p} +
     \sup_{y>0}y^{1-\alpha{}}\|D_{y}P_{y}\vf \| _{\infty{}}, \quad D_x:=\dst\frac d{dy}\\
&\|\vf\|_{\bZ^\al_p}=\|\vf\|_{p}+\sup_{y>0}y^{1-\alpha{}}\|
     D_xP_{y}\vf \| _{\infty{}}, \hspace{1cm} k=1,2,...n.
\end{eqnarray*}
define equivalent norms in the space $\bZ_p^\alpha(\bR)$.
\end{lemma}
The proof of the formulated Lemma for the case $p=\infty$ is exposed in \cite[\S\, V.4]{St70}, while for $1<p<\infty$ in \cite{DS93}.   \QED

The foregoing lemma leads to the following definition of the space $\bZ_p^\alpha(\bR)$ for $0<\alpha\leqslant\infty$, $1\leqslant p\leqslant\infty$:
\begin{eqnarray}\label{e24a}
\bZ_p^\alpha(\bR)&=&\Big\{\vf \in \bL_p(\bR)\; :\;
     \|\vf\|_{\bZ_p^{\al}}=\|\vf\|_p\nonumber\\[2mm]
&+&\sup_{y>0}y^{k-\alpha{}}\| D_{y}^{k}W^0_{a_y}\vf\|
     _{\infty{}}<\infty,\qquad k=[\alpha{}]+1\Big\},
\end{eqnarray}
where $[\alpha]$ denotes the integer part of $\alpha.$

Moreover, the definition \eqref{e24a} can be applied for the definition of the H\"older-Zygmound space $\bZ_p^\alpha(G)$ on the Lie group:
\begin{eqnarray*}
\bZ_p^\alpha(G)&=&\Big\{\psi \in \bL_p(G,dG(x))\; :\; \|\psi\|_{
    \bZ_p^{\al}}=\|\vf\|_p\nonumber\\[2mm]
&+&\sup_{y>0}y^{k-\alpha{}}\| \fD_{y}^{k}W^0_{G,a_y}\vf\|
     _{\infty{}}<\infty,\qquad k=[\alpha{}]+1\Big\}
\end{eqnarray*}
as the image of the space $\bZ_p^\alpha(\bR)$ under the transformation $t_*$ (cf. \eqref{e1.5}): $\bZ^\al_p(G):=t_*\bZ^\al_p(\bR):=\left\{\vf_0=t_*\vf\;:\;\vf\in
\bZ^\al_p(\bR)\right\}$.

Further properties of the space $\bZ^\al_p(\bR)$ (valid also for $\bZ^\al_p(G)$) one can found in \cite{DS93,St70}.

In the next Lemma \ref{l2.1} we have collected some formulae on $\cF\hskip-1mm_G$-transformations. Most of them can also be found in \cite{Pe06a}.
 %
\begin{lemma}\label{l2.1}
The following holds:
\begin{subequations}
\begin{eqnarray}\label{e1.6a}
\partial_t \varphi(t)=\partial_t(x_*\varphi_0)(t)&=&x_*\fD_G
     \varphi_0(t),\quad\varphi_0=t_*\varphi\in\bC_0^\infty(G),\quad
     t\in\bR,\\[3mm]
\label{e1.6b}
\pa_t =W^0_{-i\xi},\qquad\fD_G&=&(1-x^2)\pa_x
     =W^0_{G,-2i\xi},\\[3mm]
\label{e1.6c}
&&\hskip-33mm(\cF\hskip-1mm_G f)(\xi)=-\frac1{2i\xi}\int_{-1}^1
     \left(\frac{1+y}{1-y}\right)^{i\xi}f'(y)dy,\quad
     f\in\bC_0^\infty(G),\quad\xi\in\bR,\\[3mm]
\label{e1.6d}
(\cF\hskip-1mm_G\fD_G\varphi_0)(\xi)&=&-2i\xi(\cF\hskip-1mm_G
     \varphi_0)(\xi),\qquad \xi\in\bR,
\end{eqnarray}
\begin{eqnarray}\label{e1.6e}
\left(\cF\hskip-1mm_G\frac1y\right)(\xi)&=&\pi i\coth(\pi\xi),
     \qquad \xi\in\bR,\\[3mm]
\label{e1.6f}
\left(\cF\hskip-1mm_G\sqrt{1-y^2}\right)(\xi)&=&\frac{\pi}{\cosh(\pi\xi)},
     \qquad \xi\in\bR,
\end{eqnarray}
\begin{eqnarray}\label{e1.6g}
\left(\cF\hskip-1mm_G\frac{\sqrt{1-y^2}}y\right)(\xi)&=&\pi i\tanh(\pi\xi),\qquad \xi\in\bR,
\end{eqnarray}
\begin{eqnarray}\label{e1.6h}
(\cF\hskip-1mm_G y)(\xi)&=&\frac{\pi i}{\sinh(\pi\xi)},\qquad \xi\in\bR,\\[3mm]
\label{e1.6i}
(\cF\hskip-1mm_G(1-y^2))(\xi)&=&\frac{2\pi\xi}{\sinh\pi\xi},\qquad \xi\in\bR.
\end{eqnarray}
Note, that the integrals in \eqref{e1.6e} and in \eqref{e1.6g} are understood in the Cauchy mean value sense.
\end{subequations}
\end{lemma}
\noindent
{\bf Proof:} We have:
\begin{eqnarray*}
&&t _*\partial_t x_*\varphi_0(x)=(t _*\partial_t \varphi_0(\tanh
     t))(x)=(t _*(\pa_x\varphi)_0(\tanh t)\pa_t\tanh t)(x)\\
&&\hskip15mm=\pa_x\varphi_0(x)t _*\left(\dst\frac{1}{\cosh^2t }
     \right)(x)=\pa_x\varphi_0(x)(t _*(1-\tanh^2t ))(x)\\
&&\hskip15mm=\pa_x\varphi_0(x)(1-x^2)=(1-x^2)\pa_x\varphi_0(x)
=\fD_G\varphi_0(x),\quad x\in G.
\end{eqnarray*}

Formulae \eqref{e1.6a} and \eqref{e1.6b} are consequences of the proved formula.

Formulae \eqref{e1.6c} follows from  \eqref{e1.6d}. Indeed,
\begin{eqnarray*}
-\frac1{2i\xi}\int_{-1}^1\left(\frac{1+y}{1-y}\right)^{i\xi}f'(y)dy&=&
     -\frac1{2i\xi}\int_{-1}^1\left(\frac{1+y}{1-y}\right)^{i\xi}(\fD_G f)(y)\frac{dy}{1-y^2}\\
&=&-\frac1{2i\xi}(\cF\hskip-1mm_G\fD_G f)(\xi)=(\cF\hskip-1mm_G f)(\xi).
\end{eqnarray*}

o prove \eqref{e1.6e} we use equality \eqref{e1.8}, formula (cf. \cite[3.511.5]{GR07})
 \begin{equation}\label{e2.5}
\int_0^\infty \frac{\sinh(at)\cosh(bt)\,dt}{\sinh\,t}
     =\frac\pi2\frac{\sin(a\pi)}{\cos(a\pi)+\cos(b\pi)}, \qquad \qquad {\rm Re}(a+b)<1
 \end{equation}
and proceed as follows:
\begin{eqnarray*}
\left(\cF_G\frac1y\right)(\xi)&=&\left(\cF_\bR t _*^{-1}\frac1y
     \right)(2\xi)=\int_{-\infty}^\infty e^{i2\xi t}\coth\,t\,dt\nonumber\\
&=&2\int_0^\infty \frac{\sinh(2\xi ti)\cosh\,t\,dt}{\sinh\,t}
     =-\pi\frac{\sin(2\pi\xi i)}{1-\cos(2\pi\xi i)}\nonumber\\
&=&-\pi i\frac{\sinh(2\pi\xi)}{1-\cosh(2\pi\xi)}=\pi i\frac{
     \sinh(\pi\xi)\cosh(\pi\xi)}{\sinh^2(\pi\xi)}\nonumber\\
&=&\pi i\coth(\pi\xi),\qquad \xi\in\bR.
\end{eqnarray*}

To prove \eqref{e1.6f} we use equality \eqref{e1.8}, formula (cf. \cite[3.511.4]{GR07})
 \[
\int_0^\infty \frac{\cosh(at)\,dt}{\cosh(bt)}
     =\frac\pi{2b}\sec\frac{a\pi}{2b},\qquad {\rm Re}\,b>{\rm Re}\,a
 \]
and proceed as follows:
\begin{eqnarray*}
\left(\cF_G\sqrt{1-y^2}\right)(\xi)&=&\left(\cF_\bR t _*^{-1}
     \sqrt{1-y^2}\right)(2\xi)=\int_{-\infty}^\infty e^{2t\xi i}
     \sqrt{1-\tanh^2t}dt\nonumber\\
&=&2\int_0^\infty \frac{\cosh(2\xi ti)dt}{\cosh\,t}
     =\frac\pi{\cos(\pi\xi i)}=\frac\pi{\cosh(\pi\xi)},\qquad \xi\in\bR.
\end{eqnarray*}

To prove \eqref{e1.6g} we use equality \eqref{e1.8}, formula \eqref{e2.5} and proceed as follows:
\begin{eqnarray*}
\left(\cF_G\frac{\sqrt{1-y^2}}y\right)(\xi)&=&\left(\cF_\bR
     t_*^{-1}\frac{\sqrt{1-y^2}}y\right)(2\xi)=\int_{-\infty}^\infty e^{2t\xi i}\coth\,t\sqrt{1-\tanh^2t}dt\\
&=&2\int_0^\infty \frac{\sinh(2\xi ti)\,dt}{\sinh\,t}
     =\pi\frac{\sin(2\pi\xi i)}{\cos(2\pi\xi i)+1}\\
&=&\pi i\frac{\sinh(2\pi\xi)}{\cosh(2\pi\xi)+1}
     =\pi i\frac{\sinh(\pi\xi)\cosh(\pi\xi)}{\cosh^2(\pi\xi)}\\
&=&\pi i\tanh(\pi\xi),\qquad \xi\in\bR.
\end{eqnarray*}

To prove \eqref{e1.6h} we apply, consequtively, formula \eqref{e1.8}, replace the variable under the integral $y=-\tanh\,t $, apply formula (cf. \cite[3.511.7]{GR07})
\begin{eqnarray*}
\int_0^\infty \frac{\sinh(at )\sinh(bt )dt }{\cosh(ct )}
     =\frac\pi c\frac{\sin\dst\frac{a\pi}{2c}\sin\dst\frac{b\pi}{2c}}{\cos\dst\frac{a\pi}c +\cos\dst\frac{b\pi}c}\qquad {\rm Re}\,a+{\rm Re}\,b<{\rm Re}\,c
\end{eqnarray*}
and proceed as follows:
\begin{eqnarray*}
\left(\cF_G y\right)(\xi)&=&\left(\cF_\bR(x_*y)\right)(2\xi)
     =\int_{-\infty}^\infty e^{i2\xi t}\tanh\,t\,dt\nonumber\\
&=&2\int_0^\infty \frac{\sinh(2\xi ti)\sinh\,t\,dt}{\cosh\,t}
     =2\pi\frac{\sin(\pi\xi i)\sin\dst\frac\pi2}{\cos(2\pi\xi i)+\cos\pi}\nonumber\\
&=&2\pi i\frac{\sinh(\pi\xi)}{\cosh(2\pi\xi)-1}
     =\frac{\pi i}{\sinh(\pi\xi)},\qquad \xi\in\bR.
\end{eqnarray*}

To prove the last formula \eqref{e1.6i} we change the variable under the integral $y=\tanh\,t$
\begin{eqnarray*}
(\cF_G(1-y^2))(\xi)&=&\int_{-1}^1\left(\frac{1+y}{1-y}\right)^{i\xi}
     (1-y^2)\frac{dy}{1-y^2}=\int_{-\infty}^\infty
     e^{2t\xi i }(1-\tanh^2t)dt\\
&=&\int_{-\infty}^\infty e^{2t\xi i}\frac{dt}{\cosh^2t}
=2\int_0^\infty\frac{\cos(2\xi t)}{\cosh^2t}dt
\end{eqnarray*}
and by applying formula (cf. \cite[3.982.1]{GR07})
\begin{eqnarray*}
\int_0^\infty \frac{\cos(at)dt}{\cosh^2\beta t}=\frac{a\pi}{2\beta^2
     \sinh\dst\frac{a\pi}{2\beta}},\qquad {\rm Re}\,\beta>0,\quad a>0.
\end{eqnarray*}
we immediately derive \eqref{e1.6i}.

Formulae \eqref{e1.6a}-\eqref{e1.6i} are proved.    \QED

Note that according to \eqref{e1.6a} and \eqref{e1.6b} the counterpart of the derivative  $\pa_t $ on the Lie group $\bR$ is the derivative $\fD_G:=(1-x^2)\partial_x$ on the Lie group $G $ and both of them are convolution operators.

The notation $\bW^m_p(G,dG(x))$ with $1\leqslant p\leqslant\infty,\; m\in \bN_0$ refers to the Sobolev space of functions which represents the closure of the Schwartz space $\bS(G)$ under the norm
 \[
\| f\,|\, \bW_p^m(G,dG(x))\|\,:=\,\left(\sum\limits_{k=0}^m\|\fD^k_G f\,|\,
    \bL_p(G,dG(x))\|^p\right)^{1/p}
 \]
for $1\leqslant p<\infty$, with the usual ${\rm ess}\,\sup$-norm modification for $p=\infty$:
 \[
\|f\,|\, \bW_\infty^m(G)\|\,:=\,\sum\limits_{k=0}^m{\rm ess}\,\sup\limits_{x\in G}|\fD^k_G f(x)|\, .
 \]
The functions from $\bW^m_p(G,dG(x))\subset\bL_p(G,dG(x))$ have distributional derivatives $\fD_G^k\varphi$ of order $k=1,\ldots,m$, which belong to the Lebesgue space $\bL_p(G,dG(x))$.

The notation $\bH^s_p(G ))=\bH^s_p(G,dG(x))$  with $s\in\bR,\;1\leqslant p\leqslant\infty$ refers to the Bessel potential space on the Lie group $\bR$, where the norm is defined as follows
 \begin{eqnarray}\label{e2.6}
\|f\,|\,\bH_p^s(G,dG(x))\| \,:=\, \| {\cF}_G^{-1} \langle\xi\rangle^s \cF\hskip-1mm_G f \,|\,
     \bL_p(G,dG(x))\|\nonumber\\
=\, \| W^0_{\la\cdot\ra^s}f\,|\, \bL_p(G,dG(x))\|<+\infty, \qquad
     \langle\xi\rangle^s:=(1+|\xi|^2)^{s/2}
 \end{eqnarray}
for $1\leqslant p\leqslant\infty$. We will apply the standard convention and write $\bH^s(G,dG(x))$ for the Hilbert space $\bH^s_2(G,dG(x))$, by dropping the subscript index $p=2$. Due to the isomorphisms \eqref{e1.10} the following
 \begin{eqnarray*}
\|f\,|\,\bH^s(G,dG(x))\|_0:=\left(\int_{-1}^1\left|(1+\xi^2)^{s/2}\cF\hskip-1mm_G\,
    f(\xi)\right|^2\,d\xi\right)^{1/2}
 \end{eqnarray*}
defines an equivalent norm on $\bH^s(\bR)$.

Let $\cL(\fB_1,\fB_2)$ denote the space of all linear bounded operators mapping the Banach spaces $\fB_1\longrightarrow\fB_2$.

Based on formulae \eqref{e2.3} and \eqref{e1.6d} the following is proved (cf. \cite{Du79,Tr95} for the case $\bH_p^m(\bR)=\bW_p^m(\bR)$).
 %
\begin{proposition}\label{p2.2} For arbitrary $1<p<\infty$ and an integer $m\in\bN_0$ the the Bessel potential space $\bH^m_p(G,dG(x))$ and the Sobolev space $\bW^m_p(G,dG(x))$ have equivalent norms and are topologically isomorphic.
 \end{proposition}
Due to the isomorphisms
\begin{eqnarray}\label{e2.8}
\begin{array}{rcl}
  t _*&:&\bH_p^s(\bR)\longrightarrow\bH^s_p(G,dG(x)),\qquad 1\leqslant p\leqslant\infty,\quad s\in\bR,\\[3mm]
  t _*&:&\bZ^\mu_p(\bR)\longrightarrow\bZ^\mu_p(G),\qquad 0<\mu<\infty,\quad 1\leqslant p\leqslant\infty,
\end{array}
 \end{eqnarray}
we can justify the following propositions, valid for the Bessel potential and H\"older-Zygmound spaces on the Euclidean space $\bR$.
 %
\begin{proposition}(see \cite[\S\,2.4.2]{Tr95})\label{p2.3}
Let $s_0,s_1,r_0,r_1\in\bR, \quad 1\leqslant p_0,p_1,q_0,q_1<\infty,$ $0<\ t  <1$ and
\[
\begin{array}{c}
\dst\frac1p= \frac{1-\theta  }{p_0}+ \frac{\theta}{p_1},  \quad \dst\frac1q=\frac{1-\theta }{q_0}+ \frac{\theta}{q_1},\quad s=(1-\theta)s_0+\theta s_1\, ,\quad r=(1-\theta)r_0+\theta r_1\, .
\end{array}
\]

If $A\in\cL_j:=\cL(\bH^{s_j}_{p_j}(G,dG(x)),\bH^{r_j}_{q_j}(G,dG(x)))$, $j=0,1$, then $A$ is bounded between the interpolated spaces $A\in\cL:=\cL(\bH^s_p(G,dG(x)),\bH^r_q(G,dG(x)))$ and the norm is estimated as follows
 \[
 \|A\big|\cL\|\leqslant\|A\big|\cL_0\|^{1-\theta}\|A\big|\cL_1\|^\theta.
 \]
 \end{proposition}
 %
\begin{proposition}(see \cite[\S\,2.7.2]{Tr95})\label{p2.4}  Let $0<\mu_0,\mu_1,\nu_0,\nu_1<\infty$, $0<\gamma,\theta <1$ and
\[
\mu=(1-\theta)\mu_0+\theta\mu_1\, ,\quad \nu=(1-\theta)\nu_0+\theta\nu_1\, .
\]

If $A\in\cL_j:=\cL(\bZ^{\mu_j}_p(G),\bZ^{\nu_j}_p(G)$, $j=0,1$, then $A$ is bounded between the interpolated spaces $A\in\cL:=\cL(\bZ^\mu_p(G),\bZ^\nu_p(G)$ and the norm is estimated as follows
 \[
 \|A\big|\cL\|\leqslant\|A\big|\cL_0\|^{1-\theta}\|A\big|\cL_1\|^\theta.
 \]
 \end{proposition}
 %
\begin{proposition}[see \cite{Tr95,DS93}]\label{p2.5} Let $s,r\in\bR$, $1\leqslant p\leqslant\infty$, $0<\mu<\infty$ and $\langle\xi\rangle^r:=(1+|\xi|^2)^{1/2}$. The Bessel potential operator
 \begin{eqnarray}\label{e2.9}
 \begin{array}{rcl}
\Lbb^r:=W^0_{G,\langle\cdot\rangle^r}&:&\bH_p^s(G,dG(x))\rightarrow
     \bH_p^{s-r}(G),\\[2mm]
&:&\bZ_p^\mu(G)\rightarrow\bZ_p^{\mu-r}(G).
\end{array}
\nonumber
 \end{eqnarray}
is an isometric isomorphism for the firs pair of spaces and is an isomorphism for the second pair of spaces.
\end{proposition}

\section{\bf Convolutions and multipliers on $G$}
\label{s3}

Now we consider a convolution on the Lie group $G=(-1,1)$
 \begin{eqnarray}\label{e3.1}
(\Kb_0\varphi)(x)=(k_0*_G\varphi_0)(x):=\int_{-1}^1k_0(x\circ
    (-y))\varphi_0(y)dG(y)\nonumber\\
=\int_{-1}^1k_0\left(\frac{x-y}{1-xy}\right)\varphi_0(y)\frac{dy}{1-y^2}.
      \nonumber
 \end{eqnarray}
Convolution \eqref{e3.1} is equivalent with the corresponding convolution on the Lie group $\mathbb{R}$ (the Fourier convolution)
\begin{eqnarray}\label{e3.2}
(\Kb\varphi)(t)=(k*_\bR\varphi)(t)=\int_{-\infty}^{\infty} k(t - \tau)\varphi(\tau)\,d\tau
\end{eqnarray}
and the equivalence is established with the help of isomorphisms $t_*$ and $x_*$ (cf. \eqref{e1.5}. Indeed,
\begin{eqnarray}\label{e3.3}
t_*\K (x_*\varphi_0)(x)&=&\int_{-\infty}^{\infty} k\left(\dst
     \frac12\ln\frac{1+x}{1-x}- t \right)(x_*\varphi_0)(t) dt\nonumber\\[3mm]
&=&\int_{-1}^{1} k\left(\dst\frac12\ln\frac{1+x}{1-x}-\dst\frac12
     \ln\frac{1+y}{1-y}\right)\varphi_0(y)dG(y)\nonumber\\[3mm]
&=&\int_{-1}^1 k\left(\dst\frac12\ln\dst\frac{1+\dst\frac{x-y}{1-
     xy}}{1-\dst\frac{x-y}{1-xy}}\right)\varphi_0(y)dG(y)\nonumber\\[3mm]
&=&\int_{-1}^1 k_0\left(\dst\frac{x-y}{1-xy}\right)\,
     \varphi_0(y)dG(y)=(\Kb_0\varphi_0)(x),\quad x\in G,
\end{eqnarray}
where we have inserted $ t =\dst\frac12\ln\frac{1+y}{1-y}$. Vice versa, we start from the convolution \eqref{e3.1}:
\begin{eqnarray}\label{e3.4}
&& x_*\K_0(t _*\varphi)(t)=\int_{-1}^{1}k_0\left(\frac{\tanh t -y}{1-y\tanh t }\right)(t _*\varphi)(y)
     dG(y)\nonumber\\
&&\hskip20mm=\int_{-\infty}^\infty k_0\left(\dst\frac{\tanh t -\tanh\tau }{1-\tanh\tau \tanh t }
     \right)\varphi(\tau)d\tau \nonumber\\
&&\hskip20mm=\int_{-\infty}^\infty k_0\left(\tanh(t-\tau)\right)\varphi(\tau)d\tau
     =(\Kb\varphi)(t),\quad t\in\bR,
\end{eqnarray}
where $t_*, x_*=t^{-1}_*$ are defined in \eqref{e1.5} and we have inserted $y=\tanh\tau$. Hence, the kernels of convolution operators $k_0(x)$ in \eqref{e3.1} and $k(t)$  in \eqref{e3.2} are connected as follows:
\begin{eqnarray}\label{e3.5}
k_0(x):=(t_*k)(x),\quad k(t)=(x_*k_0)(t),
     \quad x\in G,\quad t \in\bR.
\end{eqnarray}

The connection \eqref{e3.3} and \eqref{e3.4} between the convolution operators \eqref{e3.1} and \eqref{e3.2} was noted in \cite{Pe06a} and in \cite{SP20} and used to investigate Prandtl equation, Tricomi equation (see \cite{Pe06a,Pe06b}) and Prandtl equation in the Bessel potential spaces $\wt{\bH}^s(G,dG(x))$, $0\leqslant s\leqslant1$ (see \cite{SP20}). These equations are addressed in  \S\S\, 4-5 below.
 %
\begin{theorem}\label{t3.1}
Let $1\leqslant p\leqslant\infty$, $s\in\bR$, $\mu>0$. The convolution operator ${\bf K_0}$ on the Lie Group $G $ (cf. \eqref{e3.1}) is bounded in the Lebesgue and space
\begin{eqnarray}\label{e3.6}
 \begin{array}{rcl}
{\bf K_0}&:&\bL_p(G,dG(x))\to\bL_p(G,dG(x))\\[2mm]
         &:&\bZ_p^\mu(G)\rightarrow\bZ_p^\mu(G).
 \end{array}
\end{eqnarray}
provided its kernel $k_0(x)$ satisfies the condition $k_0\in\bL_1(G,dG(x))$, i.e.,
 \[
\|k_0\big|\bL_1(G,dG(x)\|=\int_{-1}^1|k_0(y)|dG(y)<\infty.
 \]
and
\begin{eqnarray}\label{e3.7}
\|\K_0\varphi_0\big|\bL_p(G,dG(x))\|\leqslant\|k_0\big|\bL_1(G,dG(x))\|
     \|\varphi_0\big|\bL_p(G,dG(x))\|.
\end{eqnarray}
\end{theorem}
\noindent
{\bf Proof:} For the proof in case of H\"older-Zygmound space $\bZ_p^\mu(G)$ we quote \cite[Theorem 4.12]{DS93}.

Fourier convolution operator ${\bf K}$ on the Lie Group $\bR$ (cf. \eqref{e3.2}) is bounded in the space $\bL_p(\bR)$ for a kernel $k\in\bL_1(\bR)$. Then $\K_0=W^0_{G,\cK}$, $\cK:=(\cF k_0)(\xi)$, $\xi\in\bR$ it is bounded in $\bL_p(G,dG(x))$ (cf. \eqref{e3.5}, \eqref{e1.8}). Due to Proposition \ref{p3.4} proved below.

The direct proof for the space $\bL_p(G,dG(x))$ goes as follows:
\begin{eqnarray*}
&&\|\K_0\varphi_0\big|\mathbb{L}_p(G, dG(x))\|=\left[\int_{-1}^{1}\left|\int_{-1}^1k_0
     \left(\frac{x-y}{1-xy}\right)\frac{\varphi_0(y)}{1-y^2}dy\right|^p\frac{dx}{1-x^2}\right]^{1/p}\\[3mm]
&&\hskip15mm\leqslant\Bigg[\int_{-1}^{1}\left\{\int_{-1}^1\left[\left|k_0\left(\frac{x-y}{1-xy}
     \right)\right|^{1/p}\frac{|\varphi_0(y)|}{(1-y^2)^{1/p}}\right]\right.\\[3mm]
&&\hskip25mm\left.\left[\left|k_0\left(\frac{x-y}{1-xy}\right)\right|^{1/p'}
     \frac1{(1-y^2)^{1/p'}}\right]dy\right\}^p\frac{dx}{1-x^2}\Bigg]^{1/p}\\[3mm]
&&\hskip15mm\leqslant\Bigg[\int_{-1}^{1}\int_{-1}^1\left|k_0\left(\frac{x-y}{1-xy}
     \right)\right|\frac{|\varphi_0(y)|^pdydx}{(1-x^2)(1-y^2)}\\[3mm]
&&\hskip25mm\left\{\int_{-1}^1\left|k_0\left(\frac{x-y}{1-xy}\right)\right|\frac{dy}{1-y^2}
     \right\}^{p/p'}\Bigg]^{1/p}
\end{eqnarray*}
\begin{eqnarray*}
&&\hskip15mm=\Bigg[\int_{-1}^{1}\frac{|\varphi_0(y)|^pdy}{1-y^2}\int_{-1}^1\left|k_0\left(\frac{x-y}{1-xy}
     \right)\right|\frac{dx}{1-x^2}\\[3mm]
&&\hskip25mm\left\{\int_{-1}^1\left|k_0\left(\frac{x-y}{1-xy}\right)
     \right|\frac{dy}{1-y^2}\right\}^{p/p'}\Bigg]^{1/p}.
\end{eqnarray*}
In the integrals we change the variables, using the invariance of the Haar measure with respect to the group operation (cf. \eqref{e1.2}):
\begin{eqnarray*}
\mbox{if}\quad t=\dst\frac{x-y}{1-xy}, \quad\mbox{then}\quad dG(t)=dG(x)=-dG(y).
\end{eqnarray*}
and obtain
\begin{eqnarray*}
&&\|\K_0\varphi_0\big|\mathbb{L}_p(G, dG(x))\|\\
&&\hskip15mm\leqslant\left[\int_{-1}^{1}\frac{|\varphi_0(y)|^pdy}{1-y^2}\int_{-1}^1
     \frac{|k_0(t)|dt}{1-t^2}\left\{-\int_1^{-1}\frac{|k_0(t)|dt}{1-t^2}\right\}^{p/p'}\right]^{1/p}\\[3mm]
&&\hskip15mm=\|k_0\big|\bL_1(G,dG(x)\|\|\varphi_0\big|\bL_p(G,dG(x)\|
\end{eqnarray*}
and \eqref{e3.7} is proved.   \QED

Note, that
\begin{eqnarray}
\label{e3.9}
\mathcal{F_\bR}[k\ast_\bR\varphi](\xi)=\cK(\xi)\Phi(\xi), \quad
     \cK(\xi)=\mathcal{F_\bR}k(\xi),\quad \Phi(\xi)=\mathcal{F_\bR}\varphi(\xi),\\
\label{e3.8}
\cF\hskip-1mm_G[k_0\ast_G\varphi_0](\xi)=\cK_0(\xi)\Phi_0(\xi), \quad
     \cK_0(\xi)=\cF\hskip-1mm_Gk_0(\xi),\nonumber\\
     \Phi_0(\xi)=\cF\hskip-1mm_G\varphi_0(\xi), \quad \xi\in\bR
 \end{eqnarray}
and $\cK_0(\xi)$, $\cK(\xi)$ are called the symbols of the corresponding convolution operators ${\bf K_0}$ in \eqref{e3.1}, ${\bf K}$ in \eqref{e3.2}. These symbols differ by constant dilation of the argument $\cK_0(\xi)=\cK(2\xi)$. Indeed, from \eqref{e1.8} and \eqref{e3.9} follows:
\begin{eqnarray}\label{e3.10}
\cK(2\xi)=(\mathcal{F_\bR}k)(2\xi)=\cF_\bR(x_*k_0)(2\xi)=\int_{-\infty}^\infty
     e^{2t\xi i}k_0(\tanh\, t )d t \nonumber\\
=\int_{-1}^1\left(\frac{1+y}{1-y}\right)^{i\xi}k_0(y)dG(y)=\cK_0(\xi),\qquad \xi\in\bR.
\end{eqnarray}

It is well known and follows from \eqref{e3.9} that the Fourier convolution operator ${\bf K}$ in \eqref{e3.2} is written in the form (see, e.g., \cite{Du79})
\begin{eqnarray}\label{e3.11}
(\Kb\varphi)(t )=(k\ast_\bR\varphi)(t )
     =\cF^{-1}_\bR\cK\cF_\bR\varphi(t )=:W^0_\cK\varphi(t ),\quad t \in\bR.
\end{eqnarray}

Similarly, due to \eqref{e3.8}, the convolution operator ${\bf K_0}$ in \eqref{e3.1} is written in the form
\begin{eqnarray}\label{e3.12}
(\Kb_0\varphi_0)(x)=(k_0\ast_G\varphi_0)(x)={\cF}_G^{-1}
     \cK_0\cF\hskip-1mm_G\varphi_0(x)=:W^0_{G,\cK_0}\varphi_0(x),\\
x\in G \nonumber
\end{eqnarray}

Let $\fM_p(\bR)=\fM(\bL_p(\bR))$, $1\leqslant p\leqslant\infty$, denote the algebra of Fourier $\bL_p(\bR)$-multipliers, consisting of functions for which the Fourier convolution operator $W^0_a$, $a\in\fM_p(\bR)$ is bounded in the Lebesgue space $W^0_a\;:\;\bL_p(\bR)\to\bL_p(\bR)$ (cf. \cite{Hr60,Du79}).

Let $\bW(\bR):=\{a(\xi)\;:\;a(\xi)=c+\cF k(\xi)=c+\cF\hskip-1mm_G k_0\;:\; k\in\bL_1,\quad k_0\in\bL_1(G,dG(x))\}$ denote the Wiener class of functions. Norm in $\bW(\bR)$ is introduced as follows:
 \[
\|a\|:=|c_0|+\|k\big|\bL_1(\bR)\|=|c_0|+\|k_0\big|\bL_1(G,dG(x))\|.
 \]

 %
\begin{theorem}\label{t3.2}
Let $1\leqslant p\leqslant\infty$. The convolution operator $W^0_{G,a}$ on the Lie Group $G $ (cf. \eqref{e3.1} and \eqref{e3.12}) is bounded in the Lebesgue space
\begin{eqnarray}\label{e3.13}
W^0_{G,a}\;:\;\bL_p(G,dG(x))\to\bL_p(G,dG(x))
\end{eqnarray}
if and only if the symbol $a(\xi)$ is a Fourier $\bL_p(\bR)$-multiplier, $a\in\fM_p(\bR)$.

In particular, $W^0_{G,a}$ is bounded in \eqref{e3.13} for all symbols $a(\xi)$ with a bounded variation $a\in \bV_1(\bR)$ and from the Wiener class $\bW(\bR)$.
\end{theorem}
\noindent
{\bf Proof:} The proof of the first part follows from the isomorphism properties \eqref{e1.7} and the equivalence of convolution operators
\begin{eqnarray}\label{e3.14}
W^0_{G,a_0}=t _* W^0_ax_*, \qquad a_0(\xi)=a(2\xi),
\end{eqnarray}
which, in its turn, follows from \eqref{e3.4}, \eqref{e3.11} and \eqref{e3.12}.

The inclusion $\bV_1(\bR)\subset\fM_p(\bR)$ was proved by S. B. Stechkin in \cite{St50} (also see \cite[Theorem 2.11]{Du79} for a proof). The inclusion $\bW(\bR)\subset\fM_p(\bR)$ for the Wiener class of functions follows from Theorem \ref{t3.1} and the first part of the present theorem.  \QED

Let us expose here the celebrated Mikhlin-H\"ormander-Lizorkin theorem on multipliers, which we formulate for the multi-variable case.
 %
 \begin{proposition}[Mikhlin-H\"ormander-Lizorkin]\label{p3.7}. Let $1<p<\infty$, $s,r\in\bR$. If a function $a(\xi)$ satisfies the estimates
 \begin{equation}\label{e3.25}
 |\xi|^\al|\pa^\al_\xi a(\xi)|\leqslant M_\al\langle\xi\rangle^r<\infty,\qquad \xi\in\bR^n
\end{equation}
for all $\al\in\bN^n_0$, then $a\in\fM^r_p(\bR^n)$ and
 \begin{eqnarray}\label{e3.26}
&\|W_{G^n,a}^0u\big|\bH^{s-r}_p(\bR^n)\|\leqslant C_p(a)\|u\big|\bH^s_p(\bR^n)\|
 \end{eqnarray}
for some constant $C_p(a)<\infty$.
 \end{proposition}
\noindent
{\bf Proof:} Due to the multi-variable analogues of Proposition \ref{p3.4} and  Theorem \ref{t3.5}, the multiplier class $\fM_p(\bR^n)$ is the same for the spaces $\bL_p(G^n,dG^n(x))$ and $\bH^s_p(G^n,dG^n(x))$ and   $\fM^r_p(\bR^n)=(1+|\xi|^2)^{r/2}\fM_p(\bR^n)$. Therefore, the proof is reduced to the case of miltipliers on the Lebesgue space. Then the formulated assertion follows from the classical Mikhlin-H\"ormander-Lizorkin theorem (cf. \cite[v. I, Theorem 7.9.5]{Hr83}).         \QED
 %
 \begin{proposition}\label{p3.7a}. Let $1<p<\infty$, $\mu>r>0$. If a function $a(\xi)$ satisfies the condition of the foregoing Proposition \ref{p3.7}, then the convolution operator $W^0_{G,a}$ is bounded in the H\"older-Zygmound space $\bZ_p^\mu(G)\to\bZ_p^{\mu-r}(G)$.
 \end{proposition}
\noindent
{\bf Proof:} For the proof we quote \cite[Theorem 4.8]{DS93}.  \QED
 %
\begin{theorem}\label{t3.2a}
Let $1\leqslant p\leqslant\infty$, $\mu>0$. If the function $a\in \bV_1(\bR)$ has the bounded variation $a\in \bV_1(\bR)$, the convolution operator $W^0_{G,a}$ is bounded in the H\"older-Zygmound space
\[
W^0_{G,a}\;:\;\bZ^\mu_p(G)\to\bZ^\mu_p(G).
\]
\end{theorem}
\noindent
{\bf Proof:} The convolution operator $W^0_{G,-\sign}$ is bounded in $\bZ^\mu_p(G)$ due to Proposition \ref{p3.7a}, since $\sign\,\xi$ and its derivative $\xi\pa_\xi\sign\,\xi=\xi\delta(\xi)=0$ are bounded functions.

Further the boundedness of convolution operator $W^0_{G,a}$, $a\in \bV_1(\bR)$ (Stechkin's theorem) is proved as in the original paper \cite{St50} (the proof of Stechkin, modified by Matsaev, is exposed also in \cite[Theorem 2.11]{Du79}).   \QED
 %
 \begin{remark}
Let $M(\bZ^\mu_p(G))$ denote the space of multipliers in the H\"older-Zygmound space $\bZ^\mu_p(G)$, namely, $a\in M(\bZ^\mu_p(G))$ if and only if $W^0_{G,a}$ is bounded in $\bZ^\mu_p(G)$.

Summarizing the exposed results we note that $M(\bZ^\mu_p(G))$ contains the Wiener class $W(\bR)$, functions of bounded variation $V_1\bR)$ and functions which satisfy the Mikhlin-H\"ormander conditions $a(\xi)$ and $\xi\pa_\xi a(\xi)$ are uniformly bounded on $\bR$.
 \end{remark}
 %
\begin{lemma}\label{l3.0}
Let
\begin{eqnarray}\label{e3.12a}
a(\xi)=c_0-ic_1\tanh(h\xi)+\frac{c_2}{\cosh(h\xi)},\qquad \xi\in\bR,
\end{eqnarray}
where $c_0,c_1,c_2\in\bC$, $h\in\bR$, are constants.

If $a(\xi)$ is elliptic, i.e., $\inf_{\xi\in\bR}|a(|\xi)|>0$, the inverse $a^{-1}(\xi)$ has a similar representation:
\begin{eqnarray}\label{e3.12b}
&&\hskip25mm a^{-1}(\xi)=d_0-d_1\tanh(h\xi)+a_{-1}(\xi),\nonumber\\[2mm]
&&\hskip-0mm a_{-1}\in C^\infty(\bR),\quad
     a_{-1}(\xi)=\cO\left(\cosh^{-1}(h\xi)\right)
     =\cO\left(e^{-|h\xi|}\right),\\[2mm]
&&d_0=\frac{(c_0+ic_1)^{-1}+(c_0-ic_1)^{-1}}2,\qquad d_1=\frac{(c_0+ic_1)^{-1}-(c_0-ic_1)^{-1}}2. \nonumber
\end{eqnarray}

In particular, if $h\not=0$ \eqref{e3.12b} implies $a_{-1}\in\bS(\bR)$.
\end{lemma}
{\bf Proof:} It is easy to check, that
\begin{eqnarray*}
&&\left[c_0-ic_1\tanh(h\xi)\right]\left[d_0-d_1\tanh(h\xi)\right]
     =c_0d_0+ic_1d_1\tanh^2(h\xi)\\
&&\hskip7mm-[c_0d_1+ic_1d_0]\tanh(h\xi)=c_0d_0+ic_1d_1\tanh^2(h\xi)\\
&&\hskip7mm=c_0d_0+ic_1d_1-ic_1d_1[1-\tanh^2(h\xi)]=1-\frac{ic_1d_1}{\cosh^2(h\xi)}
     ,\quad\xi\in\bR.
\end{eqnarray*}
therefore,
 \[
1\equiv a(\xi)a^{-1}(\xi)=1-\frac{ic_1d_1}{\cosh^2(h\xi)}
     +[c_0-ic_1\tanh(h\xi)]a_{-1}(\xi)
     +\frac{c_2a^{-1}(\xi)}{\cosh(h\xi)}
 \]
Then,
\begin{eqnarray}\label{e3.12c}
a_{-1}(\xi)=[c_0-ic_1\tanh(h\xi)]^{-1}\left[\frac{ic_1d_1}{\cosh^2(h\xi)}
     -\frac{c_2a^{-1}(\xi)}{\cosh(h\xi)}\right]\nonumber\\
=\cO\left(\cosh^{-1}(h\xi)\right)=\cO\left(e^{-|h\xi|}\right), \quad |\xi|\to\infty,
\end{eqnarray}
since, due to the ellipticity of $a(\xi)$,
\begin{eqnarray}\label{e3.12d}
\inf_{|\xi|>R}\left|c_0-ic_1\tanh(h\xi)\right|\not=0
\end{eqnarray}
for $R>0$ sufficiently large.

From \eqref{e3.12c} and \eqref{e3.12d} follows the claimed asymptotic in \eqref{e3.12a}.       \QED
 %
\begin{remark}\label{r3.3}
Cauchy singular integral operator
\begin{eqnarray}\label{e3.15}
\Sb_G u(x):=\frac1{\pi i}\int_{-1}^1\frac{u(y)dy}{y-x}
Let . =W^0_{\sigma(\Sb_G)}u(x)-\F(u),
     \qquad x\in G
\end{eqnarray}
is represented as a difference of a convolution operator on the Lie group modulo and one dimensional operator (functional)
 \[
\Sb_G u=W^0_{\sigma(\Sb_G)}u(x)-\F(u), \qquad \F(u)=\langle F,u\rangle:=\dst\frac1{\pi i}\dst\int_{-1}^1yu(y)dG(y).
 \]
Both operators in this difference are unbounded in the space $\bL_p(G,dG(x))$, $1<p<\infty$.
\end{remark}

Indeed,
 \begin{eqnarray*}
\Sb_G u(x)&=&\dst\frac1{\pi i}\int_{-1}^1\dst\frac{1-y^2}{y-x}
     u(y)dG(y)\\
&=&\dst\frac1{\pi i}\int_{-1}^1\dst\frac{1-xy}{y-x}
     u(y)dG(y)-\dst\frac1{\pi i}\int_{-1}^1yu(y)dG(y)\\
&=&\left(\frac1{\pi iy}\ast_G u\right)(x)-\F(u)
     =W^0_{\sigma(\Sb_G)}u(x)-\F(u),\\
\sigma(\Sb_G)(\xi)&=&\cF\hskip-1mm_G\left(\frac1{\pi iy}
     \right)(\xi)=\coth\,\pi\xi,\qquad \xi\in\bR.
 \end{eqnarray*}
and in the last line we used formula \eqref{e1.6e}.

That $\F$ is unbounded is easily checked.

As for the convolution operator, the symbol $\sigma(\Sb_G)(\xi)$ is unbounded and can not belong to any multiplier class $\sigma(\Sb_G)\not\in\fM_p(\bR)$, $\forall\, p\in[1,\infty]$. \QED

Obviously, $\fM_p(\bR)$ is a Banach algebra, endowed with the norm $\|a\big|\fM_p(\bR)\|:=\|W^0_{G,a}\big|\bL_p(G,dG(x))\|$, because
\begin{eqnarray}\label{e3.18}
W^0_{G,a}W^0_{G,b}=W^0_{G,ab}\qquad \forall a,b\in\fM_p(\bR).
\end{eqnarray}
 %
\begin{proposition}[see \cite{Hr60,Du79}]\label{p3.4}
Let $1<p<\infty$ and $s\in\bR$.  The multiplier class of the space $\bH_p^s(G,dG(x))$ is independent of the parameter $s\in\bR$
 \[
\fM(\bH_p^s(G,dG(x)))=\fM(\bL_p(G,dG(x)))=\fM_p(\bL_p(\bR))=\fM_p(\bR)\,.
 \]

The norm of a convolution operator $W^0_{G,a} $, $a\in\fM_p(\bR)$ in the Bessel potential space $\bH^s_p(G,dG(x))$ on the Lie group is independent of $s$:
 \[
\|W^0_{G,a} |\cL(\bH_p^s(G,dG(x)))\|=\|W^0_{G,a} |\cL(\bL_p(G,dG(x)))\|\qquad \forall s\in\bR.
 \]
 \end{proposition}

Let $\fM^r_p(\bR)$, $1<p<\infty$, $r\in\bR$, denote the class of functions
 \begin{equation}\label{e3.19}
\fM^r_p(\bR):=\left\{\langle\xi\rangle^r a(\xi)\;:\; a\in\fM_p(\bR)\right\}, \qquad
     \langle\xi\rangle^r=(1+|\xi|^2)^{r/2}
 \end{equation}
and use $\fM_p(\bR)$ for $\fM^0_p(\bR)$.
 %
\begin{theorem}\label{t3.5}
Let $1<p<\infty$, $s,r\in\bR$. The convolution operator
 \begin{equation}\label{e3.20}
W^0_{G,a}\::\:\bH_p^s(G,dG(x))\rightarrow\bH_p^{s-r}(G,dG(x))
 \end{equation}
is bounded if and only if $a\in\fM^r_p(\bR)$.

The convolution operator $W^0_{G,a}$ in \eqref{e3.20} is Fredholm if only the symbol is elliptic
\begin{eqnarray}\label{e3.21}
     \inf_{\xi\in\bR}\left|\langle\xi\rangle^{-r}a(\xi)\right|>0.
\end{eqnarray}
If the symbol $a_{-r}(\xi):=\langle\xi\rangle^{-r}a(\xi)$ has a bounded variation $a_{-r}\in \bV_1(\bR)$ or belongs to the Wiener class, $a_{-r}\in\bW(\bR)$, the ellipticity of the symbol \eqref{e3.21} is sufficient for $W^0_{G,a}$ to be invertible in the setting \eqref{e3.20} and the inverse operator is $W^0_{G,a^{-1}}$.
 \end{theorem}
 \noindent
{\bf Proof:} Due to the property \eqref{e3.18} and Proposition \ref{p3.4}, the operator $W^0_{G,a}$ in \eqref{e3.20} is equivalently lifted to the operator
\begin{eqnarray}\label{e3.22}
\Lbb^{s-r}W^0_{G,a}\Lbb^{-s}=W^0_{G,\langle\xi\rangle^{-r}a}\;:\;
     \bL_p(G,dG(x))\rightarrow\bL_p(G,dG(x)).
\end{eqnarray}

The equivalent lifting \eqref{e3.22} shows that the operator $W^0_{G,a}$ in \eqref{e3.20} and the operator $W^0_{G,\langle\xi\rangle^{-r}a}$ in \eqref{e3.22} are simultaneously bounded or not, are simultaneously Fredholm or not, are simultaneously invertible or not. Therefore, we only need to prove the assertion for $r=0$.

Let us note that $W^0_{G,a}$ is translation invariant
\begin{eqnarray}\label{e3.23}
\tau^0_y W^0_{G,a}u(x)=W^0_{G,a}\tau^0_yu(x), \quad\tau^0_yu(x):=u\left(\frac{x+y}{1+xy}
     \right),\quad x,y\in G.
\end{eqnarray}
This property can be checked directly, but it also follows from the equivalence of operators $W^0_{G,a}$ and $W^0_a$ (cf. \eqref{e3.14}) since $W^0_a$ is translation invariant
 \[
\tau_h W^0_a\varphi(t)=W^0_a\tau_h\varphi(t), \qquad\tau_h  \varphi(t):=\varphi(t+h),\qquad t,h\in\bR.
 \]

If we assume that $W^0_{G,a}\in\cL(\bL_p(G,dG(x)))$ is Fredholm, it is invertible. This follows since $W^0_{G,a}$ is translation invariant \eqref{e3.23} and is proved as for the Fourier convolution operator $W^0_a$ in \cite{Du79,Du84} (I. Simonenko proved this property first in \cite{Si65} for a singular integral operator).

If $W^0_{G,a}\in\cL(\bL_p(G,dG(x)))$ is invertible, the dual (adjoint) operator $W^0_{G,\ov{a}}\in\cL(\bL_{p'}(G,dG(x)))$, $p'=p/(p-1)$, is  also invertible. Then is invertible $W^0_{G,a}=JW^0_{G,\ov{a}}J$, where $Ju(x)=\ov{u(x)}$ is the complex conjugation. Then, by interpolation, the inverse is bounded in the space $\cL(\bL_2(G,dG(x)))$ and, therefore, $W^0_{G,a}\in\cL(\bL_2 (G,dG(x)))$  is invertible. Since the Fourier transforms $\cF^0_G=\dst\frac1{\sqrt{\pi}}\cF\hskip-1mm_G$ and $(\cF^0_G)^{-1}:=\sqrt{\pi}\cF\hskip-1mm_G^{-1}$ map the corresponding spaces
isometrically (maintaining the norms; see \eqref{e1.10}), the operator
\[
(\cF^0_G W^0_{G,a}(\cF^0_G)^{-1}U)(\xi)=a(\xi)U(\xi), \qquad
     U\in\bL_2(\bR)
\]
is also invertible in the space $\bL_p(\bR)$. But the latter represents a
multiplication operator by the symbol $a(\xi)$ in the space $\bL_2(\bR)$.
The invertibility of the multiplication operator by a function $aI$ implies the ellipticity of the function $a(\xi)$.

If $a_{-r}\in\bV_1(\bR)\cap\bW(\bR)\subset\fM_p(\bR)$ is elliptic, then $a^{-1}_{-r}\in\bV_1(\bR)\cap\bW(\bR)\subset\fM_p(\bR)$ and, due to the property \eqref{e3.22} $W^0_{G,a^{-1}}$  is the inverse operator to $W^0_{G,a}$:
 \[
W^0_{G,a^{-1}}W^0_{G,a}=W^0_{G,a}W^0_{G,a^{-1}}=W^0_{G,1}=I.
\hskip45mm  \mbox{\QED}
 \]
 %
\begin{remark}\label{r3.6}
For $p=2$ the ellipticity condition \eqref{e3.21} is necessary and sufficient for the convolution operator
\begin{eqnarray}\label{e3.24}
W^0_{G,a}\::\:\bH^s(G,dG(x))\rightarrow\bH^{s-r}(G,dG(x))
 \end{eqnarray}
with a symbol $a\in\fM_p(\bR)$ to be invertible, because $\fM_2(\bR)=\bL_\infty(\bR)$.

But for $p\not=2$ there exist multipliers $a\in\fM_p(\bR)$, which are  elliptic $|a(\xi)|\geqslant1$ and even continuous, but $a^{-1}\not\in\fM_p(\bR)$ (see \cite[Theorem 6]{Ig69}).
\end{remark}

\section{\bf Convolution integro-differential equation}

The symbol of the integro-differential equation \eqref{e0.6} is
\begin{eqnarray}\label{e4.2}
\cA(\xi):=\sum_{k=0}^m\left[c_k(-2i\xi)^k+d_k(-2i\xi)^{m_k+n_k}
     (\cF\hskip-1mm_G\cK_k)(\xi)\right], \qquad \xi\in\bR,\\
(\cF\hskip-1mm_G\cK_k)(\xi):=\int_{-1}^1\left(\frac{1+y}{1-y}\right)^{i\xi}\frac{\cK_k(y)dy}{1-y^2},
     \quad k=1,\ldots,m.\nonumber
\end{eqnarray}
 %
\begin{theorem}\label{t4.1}
Let $1<p<\infty$, $s\in\bR$, $m_k+n_k\leqslant m$, $k=1,2,\ldots,m$. The operator $\A$ in \eqref{e0.6} in the setting
\begin{eqnarray}\label{e4.3}
\A\;:\;\bH^s_p(G,dG(x))\to\bH^{s-m}_p(G,dG(x))
\end{eqnarray}
is Fredholm if and only if its symbol is elliptic:
\begin{eqnarray}\label{e4.4}
\inf_{\xi\in\bR}\left|\frac{\cA(\xi)}{(1+\xi^2)^{m/2}}\right|>0.
\end{eqnarray}
If the ellipticity condition \eqref{e4.4}  holds, the inverse operator is the following convolution (pseudodifferential) operator $\A^{-1}=W^0_{G,\cA^{-1}}$.
\end{theorem}
\noindent
{\bf Proof:} Due to Theorem \ref{t3.5} the operator $\A$ in \eqref{e0.6} is bounded in the setting \eqref{e4.3}. Due to formulae \eqref{e3.8} and \eqref{e1.6b}, \eqref{e1.6d} the following equality holds
 \[
(\cF\hskip-1mm_G\A u)(\xi)=\cA(\xi)(\cF\hskip-1mm_G u)(\xi),\qquad\mbox{\rm or}\quad
     \A={\cF}_G^{-1}\cA\cF\hskip-1mm_G=W^0_{G,\cA}.
 \]
It is clear, that $(1+\xi^2)^{-m/2}\cA(\xi)$ belongs to the Wiener algebra $W(\bR)$ and, due to the Wiener's theorem, the inverse symbol $(1+\xi^2)^{m/2}\cA^{-1}(\xi)$ also belongs to the Wiener's algebra $W(\bR)$, provided the symbol is elliptic. But then the inverse symbol is an $\bL_p$-multiplier for all $1<p<\infty$ (see Theorem \ref{t3.2}).

Therefore, due to Theorem \ref{t3.5},  $\A=W^0_{G,\cA}$ is invertible if and only if the ellipticity condition \eqref{e4.4} holds and the inverse operator is $\A^{-1}=W^0_{G,\cA^{-1}}$.  \QED

\section{\bf Prandtl equation}

 %
\begin{theorem}\label{t5.1}
Let $1<p<\infty$, $s\in\bR$ and $f_0\in\bH^{s-1}_p(G,dG(x))$, where $f_0(x):=(1-x^2)f(x)$. The Prandtl Equation \eqref{e0.3} has a unique solution $u\in\bH^s_p(G,dG(x))$ if and only if its symbol is elliptic:
\begin{eqnarray}\label{e5.1}
\inf_{\xi\in\bR}\left|\frac{ \cP(\xi)}{\sqrt{1+|\xi|^2}}\right|>0, \qquad \cP(\xi):=c_0+2c_1\xi\coth(\pi\xi).
\end{eqnarray}
If the ellipticity condition \eqref{e5.1}  holds, the solution is
\begin{eqnarray}\label{e5.2}
u(x)=(W^0_{G,\cP^{-1}}f_0)(x),\qquad f_0(x)=(1-y^2)f(x).
\end{eqnarray}

If the condition \eqref{e5.1} fails, equation \eqref{e0.3} is not even Fredholm in the following setting $f_0\in\bH^{s-1}_p (G,dG(x))$ and $u\in\bH^s_p(G,dG(x))$.

In particular, if $f\in\bH^r_p(G,dG(x))$, $r>1/p-1=-1/p'$, $p'=p/(p-1)$, then the solution, if it exists, belongs to H\"older-Zygmound space $u\in\bZ^\mu_p(G)$ for $\mu<r+1/p'$.
\end{theorem}
{\bf Proof:} Let us multiply equation \eqref{e0.3} by $1-x^2$ and  rewrite it in the following equivalent form:
\begin{eqnarray}\label{e5.3}
\Pb_0u(x)=c_0u(x)+(1-x^2)\frac{c_1}\pi\int_{-1}^1\dst\frac{u'(y)dy}{y-x}=f_0(x),
 \end{eqnarray}
where $f_0(x)=(1-x^2)f(x)$, $f_0\in\bH^{s-1}(G,dG)$.

By using the equality
 \[
\frac{1-x^2}{y-x}=\frac{1-xy}{y-x}+x,\qquad x,y\in G
 \]
and the property of a solution $u(-1)=u(+1)=0$ we give equation \eqref{e5.3} the following form:
\begin{eqnarray}\label{e5.4}
&&\hskip-7mm \Pb_0u(x)=c_0u(x)+\dst\frac{c_1}{\pi}(1-x^2)\int_{-1}^1
     \dst\frac{u'(y)dy}{y-x}=c_0u(x)+\dst\frac{c_1}{\pi}\int_{-1}^1u'(y)
     \dst\frac{1-xy}{y-x}dy\nonumber\\
&&+\dst\frac{c_1x}{\pi}\int_{-1}^1u'(y)dy=c_0u(x)+\dst\frac{c_1}{\pi}
     \int_{-1}^1(1-y^2)u'(y)\dst\frac{1-xy}{y-x}\dst\frac{dy}{1-y^2}\nonumber\\
&&=c_0u(x)+\frac{c_1}{\pi}\left[\fD_G u\ast_G\dst\frac1y\right](x)
     =f_0(x),\qquad x\in G,
 \end{eqnarray}
where $(v\ast_G w)(x)$ is the convolution (see \eqref{e3.1}) and $\fD_G$ is the derivative (see \eqref{e0.8}) on the Lie group $G $.

By applying the Fourier transformation $\cF\hskip-1mm_G$ (see \eqref{e1.4}) to the equation \eqref{e5.4} and taking into account formulae \eqref{e3.8}, \eqref{e1.6d}, \eqref{e1.6e}, we find the following:
\begin{eqnarray}\label{e5.5}
&&\hskip-7mm c_0U(\xi)+\dst\frac{c_1}\pi\cF\hskip-1mm_G(\fD_G u)(\xi)\cF\hskip-1mm_G
     \left(\dst\frac1y\right)(\xi)=\cP(\xi)U(\xi)=F_0(\xi),\nonumber\\
&&\cP(\xi):=c_0+2c_1\xi\coth(\pi\xi),\qquad U(\xi):=(\cF\hskip-1mm_G u)(\xi),\quad\xi\in\bR.
 \end{eqnarray}

Solvability (Fredholmness) of the equation \eqref{e0.3} under conditions of Theorem \ref{t5.1} means invertibility (Fredholmness) of the operator
\begin{eqnarray}\label{e5.6}
\Pb_0=W^0_{G,\cP}\;:\;\bH^s_p(G,dG(x))\to\bH^{s-1}_p(G,dG(x)).
\end{eqnarray}
It is clear, that $(1+\xi^2)^{-1/2}\cP(\xi)$ has bounded variation (belongs to $\bV_1(\bR)$) and the inverse symbol $(1+\xi^2)^{m/2}\cA^{-1}(\xi)$ also has bounded variation, provided the symbol is elliptic. But then the inverse symbol is an $\bL_p$-multiplier for all $1<p<\infty$ (see Theorem \ref{t3.2}).

From Theorem \ref{t3.5} follows that the operator $\Pb_0=W^0_{G,\cP}$ in \eqref{e5.6} is invertible if and only if the ellipticity condition \eqref{e5.1}  holds and is not Fredholm if the ellipticity condition fails.  The solution to equation \eqref{e5.3} (and to equation \eqref{e0.3}) is represented by formula \eqref{e5.2}.

The assertion about a priori smoothness of a solution to equation \eqref{e0.3} follows from the inclusion $u\in\bH^{r+1}_p(G,dG(x))$ and the Sobolev's embedding theorem (cf. \cite{Tr95} for details):
 \begin{eqnarray}\label{e5.7}
\bH^{r+1}_p(G,dG(x))\subset\bZ_p^\mu(G) \qquad {\rm for}\quad 0<\mu<r+1-\frac1p=r+1/p'.
 \end{eqnarray}
The proof is completed.  \QED

\section{\bf Tricomi equation}

 %
\begin{theorem}\label{t6.1}
Let $1<p<\infty$, $s\in\bR$, $g_0\in\bH^s_p(G,dG(x))$, where $g_0(x):=(1-x^2)^{1/2}g(x)$.  Tricomi Equation \eqref{e0.4}  has a unique solution $v(x)$, such that $v_0:=(1-x^2)^{1/2}v\in\bH^s_p(G,dG(x))$, if and only if its symbol is elliptic:
\begin{eqnarray}\label{e6.1}
\inf_{\xi\in\bR}|\cT(\xi)|>0, \qquad \cT(\xi):=c_0-ic_1\tanh(\pi\xi)+\frac{c_2}{\cosh(\pi\xi)},\quad\xi\in\bR.
\end{eqnarray}
If the ellipticity condition \eqref{e6.1}  holds, the solution is represented as follows
\begin{eqnarray}\label{e6.2}
v(x)&=&d_0g(x)-\frac{d_1}\pi\int_{-1}^1\sqrt{\frac{1-y^2}{1-x^2}}\dst\frac{g(y)dy}{y-x}
     \nonumber\\
&&+\int_{-1}^1\sqrt{\frac{1-y^2}{1-x^2}}k_\T\left(\frac{x-y}{1-xy}\right)g(y)dy,
     \qquad x\in G,
\end{eqnarray}
where
 \[
d_0=\dst\frac{(c_0+ic_1)^{-1}+(c_0-ic_1)^{-1}}2,\qquad d_1=\dst\frac{(c_0+ic_1)^{-1}-(c_0-ic_1)^{-1}}2
 \]
and $k_\T\in\bS(G)$ is the inverse Fourier transform of the inverse symbol:
\begin{eqnarray*}
k_\T(x):=({\cF}_G^{-1}\cT_{-1})(x)=\dst\frac1{\pi}\dst
     \int_{-\infty}^\infty\left(\frac{1+x}{1-x}\right)^{-i\xi}
     \cT_{-1}(\xi)d\xi,\quad x\in G,\\
\cT_{-1}(\xi)=\cT^{-1}(\xi)-d_0-d_1\tanh(\pi\xi),\quad
\cT_{-1}\in\bS(\bR).
\end{eqnarray*}

If the condition \eqref{e6.1} fails, equation \eqref{e0.4}  is not Fredholm in the space setting $v_0,g_0\in\bH^s_p(G,G(x))$, where $v_0,g_0$ are defined above.

In particular, if $g_0\in\bH^s_p(G,dG(x))$, $s>1/p$, then the solution, if it exists, belongs to weighted H\"older-Zygmound space: $v(x)=\dst\frac{v_0(x)}{\sqrt{1-x^2}}$, where $v_0\in\bZ^\mu_p(G)$ for $\mu<s-1/p$.
\end{theorem}
{\bf Proof:} Let us multiply equation \eqref{e0.4}  by $\sqrt{1-x^2}$, use the notation of  functions $v_0$ and $g_0$ introduced above, and  rewrite \eqref{e0.4} in the following equivalent form:
\begin{eqnarray}\label{e6.4}
\T_0 v_0(x)=c_0v_0(x)+\frac{c_1}\pi\int_{-1}^1\frac{\sqrt{(1-x^2)(1-y^2)}}{y-x}
     v_0(y)dG(y)\nonumber\\
+\frac{c_2}\pi\int_{-1}^1\frac{\sqrt{(1-x^2)(1-y^2)}}{1-xy}v_0(y)dG(y)\nonumber\\
=c_0v_0(x)-\frac{c_1}\pi\int_{-1}^1\frac{1-xy}{x-y}\sqrt{1-\left(
     \frac{x-y}{1-xy}\right)^2}v_0(y)dG(y)\nonumber\\
+\frac{c_2}\pi\int_{-1}^1\sqrt{1-\left(\frac{x-y}{1-xy}\right)^2}
     v_0(y)dG(y)\nonumber\\
=c_0v_0(x)-\frac{c_1}\pi(y^{-1}\sqrt{1-y^2}\ast_G v_0)(x)
     +\frac{c_2}\pi(\sqrt{1-y^2}\ast_G v_0)(x)=g_0(x),
\end{eqnarray}
where $(v\ast_G w)(x)$ is the convolution on the Lie group $G $ (see \eqref{e1.4}).

By applying the Fourier transformation $\cF\hskip-1mm_G$ (see \eqref{e1.4}) to the equation \eqref{e6.4} and taking into account formulae \eqref{e3.8}, \eqref{e1.6f}, \eqref{e1.6g}, we find the following:
\begin{eqnarray}\label{e6.5}
\cT(\xi)V_0(\xi)&=&\left[c_0-\dst\frac{c_1}\pi\left(\cF\hskip-1mm_G\frac{\sqrt{1-y^2}}y
     \right)(\xi)+\dst\frac{c_2}\pi\cF\hskip-1mm_G\left(\sqrt{1-y^2}\right)(\xi) \right]V_0(\xi)\nonumber\\
&=&G_0(\xi),\qquad \cT(\xi):=c_0-ic_1\tanh(\pi\xi)+\frac{c_2}{\cosh(\pi\xi)},\\
    &&V_0(\xi):=(\cF\hskip-1mm_G v_0)(\xi),\quad G_0(\xi):=(\cF\hskip-1mm_G g_0)(\xi),\quad\xi\in\bR.\nonumber
 \end{eqnarray}

The solvability (Fredholmness) of equation \eqref{e0.4}  under conditions of Theorem \ref{t6.1} means the invertibility (Fredholmness) of the operator
\begin{eqnarray}\label{e6.6}
\T_0=W^0_{G,\cT}\;:\;\bH^s_p(G,dG(x))\to\bH^s_p(G,dG(x)).
\end{eqnarray}
From Theorem \ref{t3.5} follows that the operator $\T_0=W^0_{G,\cT}$ in \eqref{e6.6} is invertible if and only if the ellipticity condition \eqref{e6.1} holds. The solution to equation \eqref{e6.4} is $v_0(x)=\T^{-1}_0g_0(x)=W^0_{G,\cT^{-1}}g_0(x)$ and, therefore, the solution to equation \eqref{e0.4} is
 \[
v(x)=(1-x^2)^{-1/2}(W^0_{G,\cT^{-1}}(1-y^2)^{1/2}g)(x).
 \]
From Lemma \ref{l3.0} follows, that
\begin{eqnarray*}
\cT^{-1}(\xi)=d_0-d_1\tanh(\pi\xi)+\cT_{-1}(\xi), \qquad
     k_\T(\xi)=({\cF}_G^{-1} \cT_{-1})(\xi), \\
\cT_{-1}\in\bS(\bR)\quad\mbox{\rm and, therefore,}\quad
     k_\T\in\bS(G)
\end{eqnarray*}
and formula \eqref{e6.2} is proved.

The concluding assertion about a priori smoothness of a solution to equation \eqref{e0.4}  follows from the Sobolev's embedding theorem \eqref{e5.7}.     \QED

\section{\bf Lavrentjev-Bitsadze equation}

 %
\begin{theorem}\label{t7.1}
Let $1<p<\infty$, $s\in\bR$. The Lavrentjev-Bitsadze Equation \eqref{e0.5}  has a unique solution $\varphi$ in the setting
\begin{eqnarray}\label{e7.1}
\V h,\V\varphi\in\bH^s_p(G,dG(x)),\quad \V\psi(x):=(1-x^2)\psi\left(\frac{1+x}2\right),
     \quad x\in G,
\end{eqnarray}
if and only if its symbol is elliptic:
\begin{eqnarray}\label{e7.2}
\inf_{\xi\in\bR}|\cL\cB(\xi)|>0, \qquad \cL\cB(\xi):=c_0-ic_1\tanh\frac{\pi\xi}2,\quad\xi\in\bR.
\end{eqnarray}
If the ellipticity condition \eqref{e7.2}  holds, the solution is
\begin{eqnarray}\label{e7.3}
\varphi(x)&=&d_0h(x)-\frac{d_1}{\pi i}\int_0^1\left[\dst\frac1{y
     -x}+\dst\frac{1-2y}{x+y-2xy}\right]\dst\frac{h(y)dy}{y-x}\nonumber\\
&&+2\int_0^1k_{\Lb\B}\left(\frac{x-y}{x+y-2xy}
    \right)h(y)dy,\qquad x\in G^+,
\end{eqnarray}
where
 \[
d_0=\dst\frac12\left[{(c_0+ic_1)^{-1}+(c_0-ic_1)^{-1}}\right],\qquad d_1=\dst\frac12\left[{(c_0+ic_1)^{-1}-(c_0-ic_1)^{-1}}\right]
 \]
and $k_{\Lb\B}\in\bS(G)$ is the inverse Fourier transform of the modified symbol:
\begin{eqnarray*}
k_{\Lb\B}(x):=({\cF}_G^{-1}\cL\cB_{-1})(x)=\dst\frac1{\pi}\dst
     \int_{-\infty}^\infty\left(\frac{1-x}{1+x}\right)^{-i\xi}
     \cL\cB_{-1}(\xi)d\xi,\quad x\in G^+,\\
\cL\cB_{-1}(\xi)=\cL\cB^{-1}(\xi)-d_0-d_1\tanh\frac{\pi\xi}2,\\ \cL\cB_{-1}\in C^\infty(\bR),\quad \cL\cB_{-1}(\pm\infty)=0,\quad \cL\cB'_{-1}\in\bS(\bR).
\end{eqnarray*}

If the condition \eqref{e7.2}  fails, equation \eqref{e0.5}  is not Fredholm in the  setting \eqref{e7.1}.

In particular, if $\V h\in\bH^s_p(G,dG(x))$, $s>1/p$ and the symbol is elliptic, the solution belongs to H\"older-Zygmound space $\bZ^\mu(G)$ for $\mu<s-1/p$.
\end{theorem}
\noindent
{\bf Proof:} In the equation \eqref{e0.5} we change variables, introduce new unknown functions
\begin{eqnarray}\label{e7.4}
\begin{array}{r}
t=\dst\frac{1+x}2,\quad  \tau=\dst\frac{1+y}2,\quad w(x)=\V\varphi(x),\quad h_0(x)=\V h(x),\quad x,y\in G,
\end{array}
 \end{eqnarray}
multiply both sides by $1-x^2$ and obtain:
\begin{eqnarray}\label{e7.5}
&&\hskip-10mm\Lb\B_0w(x)=c_0w(x)+\frac{c_1}\pi\int_{-1}^1\frac{1-x^2}{1-y^2}\left[\frac1{y-x}
     -\frac y{1-xy}\right]w(y)dy\nonumber\\
&&=c_0w(x)+\frac{c_1}\pi\int_{-1}^1\frac{(1-x^2)(1-y^2)}{(y-x)(1-xy)}\frac{w(y)dy}{1-y^2}\\
&&=c_0w(x)+\frac{c_1}\pi\int_{-1}^1\left[\frac{x-y}{1-xy}-\frac{1-xy}{x-y}\right]\frac{w(y)dy}{1
     -y^2}=h_0(x),\quad x\in G,\nonumber
 \end{eqnarray}
because
\begin{eqnarray}\label{e7.6}
\frac{(1-x^2)(1-y^2)}{(y-x)(1-xy)}=\frac{x-y}{1-xy}-\frac{1-xy}{x-y}.
 \end{eqnarray}

Thus, $\Lb\B_0$ is a $G$-convolution operator with the kernel $t-\dst\frac1t$:
\begin{eqnarray}\label{e7.7}
\Lb\B_0w(x)=c_0w(x)+\frac{c_1}\pi\left(t-\frac1t\right)*_G w(x), \qquad x\in G.
 \end{eqnarray}
The symbol of the operator is (cf. \eqref{e1.6e} and \eqref{e1.6h}):
\begin{eqnarray}\label{e7.8}
\cL\cB(\xi)&=& c_0+\frac{c_1}\pi(\cF\hskip-1mm_G t)(\xi)
     -\frac{c_1}\pi\left(\cF\hskip-1mm_G\frac1t\right)(\xi)\nonumber\\
&=&c_0+ic_i\left[\frac1{\sinh(\pi\xi)}-\coth(\pi\xi)\right]\nonumber\\
&=& c_0+\frac{ic_1(1-\cosh(\pi\xi))}{\sinh(\pi\xi)}
     =c_0-ic_1\tanh\frac{\pi\xi}2, \qquad \xi\in\bR.
 \end{eqnarray}

From Theorem \ref{t3.5} follows that the operator $\Lb\B_0=W^0_{G,\cT}$ in \eqref{e7.5} is invertible if and only if the ellipticity condition \eqref{e7.2} holds. The solution to equation \eqref{e7.3} is $w(x)=\Lb\B^{-1}_0h_0(x)=W^0_{G,\cL\cB^{-1}}\V h(x)$ and, therefore, solution to equation \eqref{e0.5} is
 \[
\varphi(x)=\V^{-1}w(x):=\frac{w(1-2x)}{x(1-x)}
      =(W^0_{G,\cL\cB^{-1}}\V h)(x),
 \]
Since $\cL\cB(\pm\infty)=c_0\pm ic_1$, from Lemma \ref{l3.0} we conclude
\begin{eqnarray*}
\cL\cB^{-1}(\xi)=d_0-d_1\tanh\frac{\pi\xi}2+\cL\cB_{-1}(\xi), \qquad \cL\cB_{-1}\in\bS(\bR), \\
     k_{\Lb\B}(x)=(\cF\hskip-1mm_G^{-1}\cL\cB_{-1})(x),\quad x\in G,\qquad k_{\Lb\B}\in\bS(G)
\end{eqnarray*}
and formula \eqref{e7.3} is proved, from which follows formula \eqref{e7.3} by the change of the integration variable.

The concluding assertion about a priori smoothness of a solution to equation \eqref{e0.5}  follows from the Sobolev's embedding theorem \eqref{e5.7}.     \QED

\begin{acknowledgements}
{The investigation is supported by the grant of the Shota Rustaveli Georgian National Science Foundation FR-19-676}
\end{acknowledgements}

\end{document}